\documentclass[fleqn,usenatbib]{mnras}

\usepackage{newtxtext,newtxmath}
\usepackage[T1]{fontenc}

\DeclareRobustCommand{\VAN}[3]{#2}
\let\VANthebibliography\thebibliography
\def\thebibliography{\DeclareRobustCommand{\VAN}[3]{##3}\VANthebibliography}

\usepackage{graphicx}	% Including figure files
\usepackage{amsmath}	% Advanced maths commands
\usepackage{url}
\usepackage{enumitem}
\newcommand{\vect}[1]{\mathbf{#1}}
\newcommand{\matr}[1]{\overline{\mathbf{#1}}}

\newcommand{\kext}{\kappa_{\text{ext}}}
\newcommand{\amax}{a_{\text{max}}}

\title[Scattering by Aligned Grains]{Thermal Emission and Scattering by Aligned Grains: Plane-Parallel Model and Application to Multiwavelength Polarization of the HL Tau Disk}

\author[Z.-Y. D. Lin et al.]{
Zhe-Yu Daniel Lin,$^{1}$\thanks{Jefferson Fellow. E-mail: zdl3gk@virginia.edu}
Zhi-Yun Li,$^{1}$
Haifeng Yang,$^{2,3}$
Ian Stephens,$^{4}$
Leslie Looney,$^{5}$
\newauthor
Rachel Harrison,$^{5}$
and Manuel Fern\'andez-L\'opez$^{6}$
\\
$^{1}$Department of Astronomy, University of Virginia, 530 McCormick Rd., Charlottesville 22904, Virginia, USA\\
$^{2}$Institute for Advanced Study, Tsinghua University, Beijing, 100084, People's Republic of China \\
$^{3}$Kavli Institute for Astronomy and Astrophysics, Peking University, Yi He Yuan Lu 5, Haidian Qu, Beijing 100871, People's Republic of China \\
$^{4}$Department of Earth, Environment and Physics, Worcester State University, Worcester, MA 01602, USA \\
$^{5}$Department of Astronomy, University of Illinois at Urbana-Champaign, Urbana IL 61801, USA \\
$^{6}$Instituto Argentino de Radioastronom\'ia (CCT-La Plata, CONICET; CICPBA), C.C. No. 5, 1894, Villa Elisa, Buenos Aires, Argentina
}

\date{Accepted XXX. Received YYY; in original form ZZZ}

\pubyear{2015}

\begin{document}
\label{firstpage}
\pagerange{\pageref{firstpage}--\pageref{lastpage}}
\maketitle

\begin{abstract}

% continuum polarization of disks show scattering for disks. observed for multiple wavelength => polarization spectrum
% what is the problem and why is the problem important
% what we did: plane-parallel slab, disentangle scattering and thermal polarization
% what are our results: derive the grain size from S, 
% why our results are important / implications
Telescopes are now able to resolve dust polarization across circumstellar disks at multiple wavelengths, allowing the study of the polarization spectrum. Most disks show clear evidence of dust scattering through their unidirectional polarization pattern typically at the shorter wavelength of $\sim 870 \mu$m. However, certain disks show an elliptical pattern at $\sim 3$mm, which is likely due to aligned grains. With HL Tau, its polarization pattern at $\sim 1.3$mm shows a transition between the two patterns making it the first example to reveal such transition. We use the T-matrix method to model elongated dust grains and properly treat scattering of aligned non-spherical grains with a plane-parallel slab model. We demonstrate that a change in optical depth can naturally explain the polarization transition of HL Tau. At low optical depths, the thermal polarization dominates, while at high optical depths, dichroic extinction effectively takes out the thermal polarization and scattering polarization dominates. Motivated by results from the plane-parallel slab, we develop a simple technique to disentangle thermal polarization of the aligned grains $T_{0}$ and polarization due to scattering $S$ using the azimuthal variation of the polarization fraction. We find that, with increasing wavelength, the fractional polarization spectrum of the scattering component $S$ decreases, while the thermal component $T_{0}$ increases, which is expected since the optical depth decreases. We find several other sources similar to HL Tau that can be explained by azimuthally aligned scattering prolate grains when including optical depth effects. In addition, we explore how spirally aligned grains with scattering can appear in polarization images. 

%From the analysis of $S$, we cannot find a maximum grain size cutoff in a power-law distribution under the assumption of compact spherical grains that can explain all three wavelength continuum polarization observations of HL Tau. 

\end{abstract}


%%%%%%%%%%%%%%%%%%%%%%%%%%%%%%%%%%%%%%%%%%%%%%%%%%

%%%%%%%%%%%%%%%%% BODY OF PAPER %%%%%%%%%%%%%%%%%%

\section{Introduction} \label{sec:introduction}

Dust polarization is a unique tool that provides insight to the properties of grains. With incredible sensitivity, the Atacama Large Millimeter/Submillimeter Array (ALMA) opened a new avenue of research in millimeter-wave dust polarization for protoplanetary disks offering spatially resolved images across multiple wavelengths \citep[e.g.][]{Stephens2017, Stephens2020, Harris2018, Harrison2019, Harrison2021, Alves2018, Lee2021, Ohashi2018}. 

There are two main mechanisms for producing dust polarization. Elongated grains can emit polarized thermal photons, and we can observe polarization if the grains are aligned. There are several mechanisms proposed for grain alignment, including radiative alignment through radiative alignment torques (RAT), mechanical alignment, and aerodynamic alignment \citep[e.g.][]{Andersson2015, Lazarian2007, Lazarian2007_MAT, Kataoka2019, Lazarian1995, Gold1952}. In particular, there is evidence that grains are aligned with magnetic fields in diffuse interstellar medium and protostellar envelopes, likely due to RAT. However, it is yet to be firmly established if RAT can also cause grain alignment in protoplanetary disks. Regardless of the alignment mechanism, a key observational feature to identify aligned grains is the consistent polarization direction across multiple wavelengths in some sources, such as BHB 07-11 \citep{Alves2018}. 
% also polarization with a 90 degree offset with temperature gradient effects -> but leave out because it can be confusing for the reader here

The second mechanism for producing dust polarization is through dust scattering. Thermal photons produced by grains can scatter off of other grains and become polarized even if the initial photon is unpolarized \citep{Kataoka2015}. Thus, even disks with purely spherical grains can produce polarization. Grains will efficiently scatter when the grain size is comparable to the observing wavelength. Since photons are maximally polarized when scattered by $90^{\circ}$ and the distribution of grains is largely confined in the disk midplane, the polarization angle seen across the disk is parallel to the disk minor axis \citep{Yang2016_inc}. Many disks exhibit this feature and favor the scattering interpretation of polarization \citep[e.g.][]{Cox2018, Bacciotti2018, Girart2018, Harris2018, Hull2018, Dent2019, Sadavoy2019, Stephens2020, Aso2021}. 

However, there are a few disks that currently show a scattering morphology in the shorter wavelength and a non-scattering morphology at the longer wavelength \citep{Harrison2019} with HL Tau as the first and best studied example \citep{Kataoka2017, Stephens2017} that has resolved polarization across three ALMA bands. At Band 7 ($\lambda=870\mu m$), HL Tau has polarization parallel to the disk minor axis and is attributed to scattering \citep{Kataoka2016_hltau, Yang2016_inc}. At Band 3 ($\lambda = 3.1mm$), the polarization angle forms an elliptical pattern which was particularly puzzling. The elliptical pattern was first attributed to radiative alignment of oblate grains, where the short axes of those grains are radially aligned \citep{Kataoka2017, Tazaki2017}, but a closer inspection suggests that only azimuthally aligned prolate grains could produce the elliptical pattern \citep{Yang2019}. Still, pure azimuthally aligned prolate grains could not explain the polarized intensity along the disk major axis which \cite{Yang2019} speculated could be compensated by scattering. Indeed, \cite{Mori2021} demonstrated that a superposition of polarization from scattering spherical grains and that from azimuthally aligned prolate grains can reproduce Band 3 of HL Tau which lends weight to the existence of relatively large aligned prolate grains that can scatter (sub)millimeter photons efficiently.

A crucial question emerges immediately: where is the thermal polarization from the prolate grains in the shorter wavelengths Bands 6 and 7? Since we expect the same aligned grains to produce a consistent polarization across wavelengths, the same elliptical pattern should be present, but it is clearly not the case. Interestingly, \cite{Stephens2017} could largely reproduce Band 6 using a morphological model mixing a unidirectional polarization (to mimic Band 7) and an azimuthal polarization (to mimic Band 3). This alludes to a possibility that the same elliptical pattern from pure thermal polarization indeed exists, but somehow lessens at Band 6 and perhaps further fades at Band 7 giving way to scattering. 

Explaining the multiwavelength polarization for HL Tau quantitatively is hindered by the difficulty in treating scattering of aligned grains in a consistent way. The morphological model \citep{Stephens2017} and models that directly add scattering and thermal polarization \citep{Yang2019, Mori2021} all point to the existence of scattering of aligned grains. \cite{Yang2016_oblate} analyzed scattering of aligned grains for disks though considering only single scattering which applies to the optically thin limit (see also \citealt{Kirchschlager2020}). However, to address the variation of the optical depth with both wavelength and distance from the center star, we need a complete treatment for scattering of aligned grains beyond the optically thin limit.

In this paper, we show that the transition seen in HL Tau is a natural consequence of a change in optical depth at different wavelengths. To gain physical insight into how the polarization from scattering aligned grains changes with optical depth, we start with a simple plane-parallel slab model where the scattering of aligned grains is treated with the T-matrix method. Section \ref{sec:plane_parallel} explains the slab set up, and Section \ref{sec:slab_results} presents the results. The results of the slab model are used to interpret the multiwavelength observations of HL Tau in Section \ref{sec:hl_tau} where we use the plane-parallel calculations to piece together images that can compare with observations. We provide an empirical method to separate the contributions to the observed polarization from scattering and thermal polarization. Section \ref{sec:discussion} discusses the implications of our findings for other wavelengths of HL Tau and other sources. Finally, we summarize our conclusions in Section \ref{sec:conclusion}.

\section{Plane Parallel Slab} \label{sec:plane_parallel}

Qualitatively, it is easy to understand why optical depth produces a transition between the polarization patterns dominated by scattering and direct emission. In the optically thin limit, a packet of photons emitted from a grain has a low chance of impacting another grain and the observer mostly sees the photons that were directly produced by the grains. If the grains emit intrinsically polarized photons (e.g., thermal polarization of elongated grains), then the observer mostly sees the thermal polarization. In the optically thick limit, photons produced by grains encounter other grains easily and the photons are absorbed or scattered. If a photon is scattered, the polarization state changes depending on the scattering direction. Photons undergo multiple scattering events with the polarization state constantly modified by each event before eventually escaping the system. As the packet of photons travel through a medium, different polarization states experience different levels of extinction which is called dichroic extinction. The observer thus sees many photons that experienced multiple interactions with the medium. Much of the direct emission is hidden because of dichroic extinction and results in polarization dominated by scattering.

Including scattering complicates the radiative transfer equation enormously. Monte Carlo techniques have been a powerful tool to treat scattering for the three dimensional structure of disks. However, the complete treatment of scattering including aligned grains is notoriously difficult and most of the work has been limited to spherical or randomly aligned grains (see e.g., \citealt{Dullemond2012}, \citealt{Steinacker2013}, \citealt{Baes2019}). Given that the dust in protoplanetary disks can be fairly geometrically thin \citep[e.g.,][]{Dutrey2017, Villenave2020} and especially for HL Tau \citep{Pinte2016}, we can use a plane-parallel slab to capture the essence of the problem. 

Plane-parallel slab calculations including scattering have been useful in understanding how scattering affects the Stokes $I$ of the slab. In particular, there are analytical solutions to the radiation transfer equation assuming isotropic scattering \citep{Rybicki1979, Miyake1993, Birnstiel2018, Zhu2019}. Albeit approximations, solutions to the plane-parallel model can be used to infer disk properties as a function of radius, instead of assuming power-law radial profiles for the surface density or temperature \citep{Carrasco2019, Macias2021, Sierra2021}. For this section, we describe the methodology and assumptions to develop the plane-parallel slab model of aligned grains with scattering.

\subsection{Problem Setup}
\begin{figure}
    \centering
    \includegraphics[width=0.95\columnwidth]{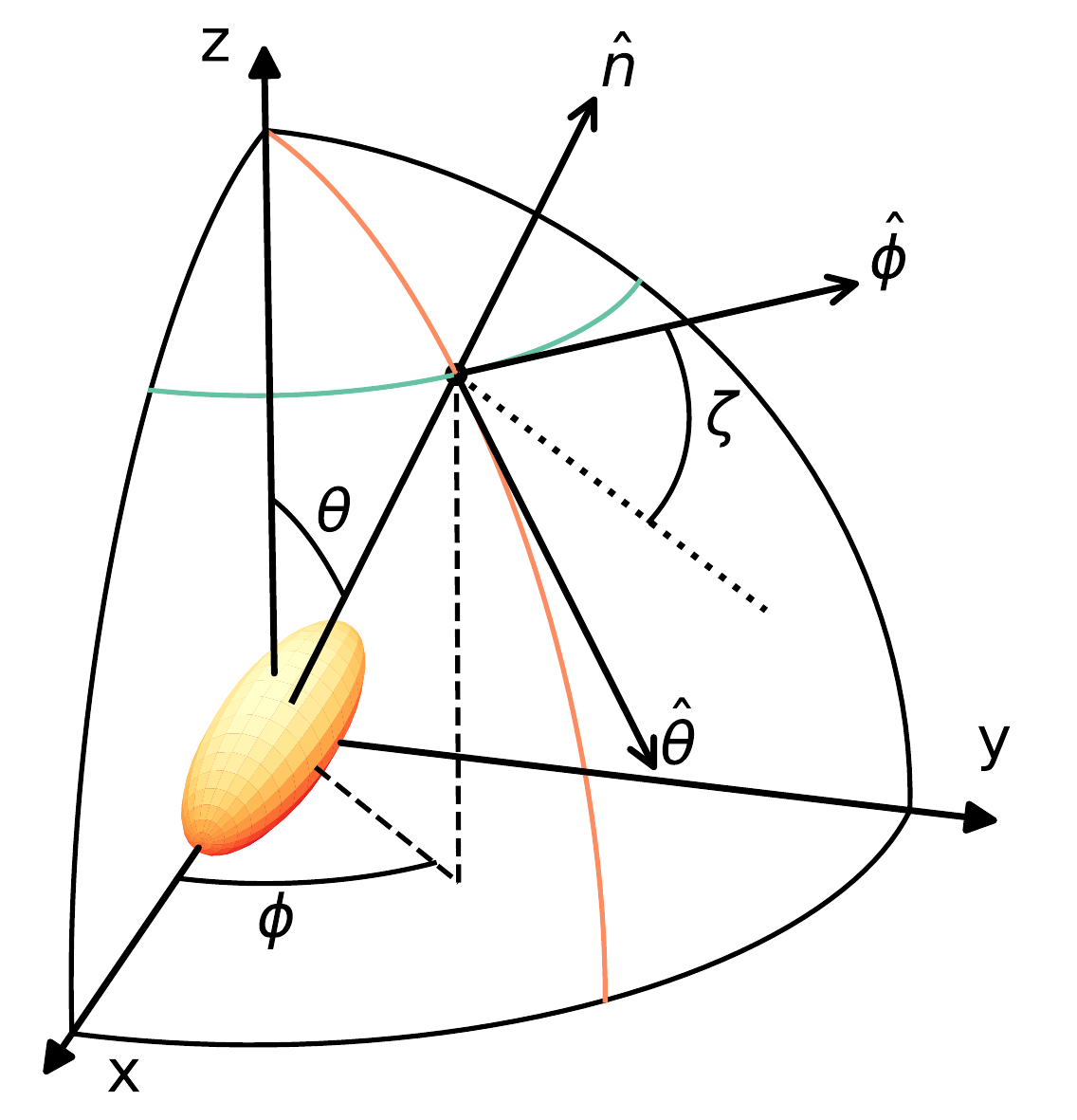}
    \caption{
        The coordinate system centered around one dust grain. The slab extends to infinity along the $x$- and $y$-axis. The orange capsule represents a prolate grain with the axis of symmetry along the $x$-axis. The line of sight is along $\hat{n}$. The direction of increasing $\theta$ and $\phi$ are $\hat{\theta}$ and $\hat{\phi}$. The orange arc is the meridian in the plane formed by $\hat{z}$ and $\hat{n}$. The green arc is of constant polar angle $\theta$. The Stokes parameters are defined on the image plane formed by $\hat{\theta}$ and $\hat{\phi}$. The polarization angle is defined by $\zeta$ which is an angle from $\hat{\phi}$ in the image plane. 
    }
    \label{fig:coordinates}
\end{figure}

Consider a slab that is infinite along the $x$- and $y$-axis direction in a Cartesian coordinate system, but finite in $z$. The slab is put between $z=0$ and $z=\Delta z$ with arbitrary units since the radiative transfer depends only on the optical depth (defined in Section~\ref{ssec:grain_model}) and not the physical depth. The slab has a uniform density $\rho$ and is isothermal with a temperature $T$. As shown in Fig. \ref{fig:coordinates}, $\theta$ is the angle from the $z$-axis and $\phi$ is the azimuthal angle from the $x$-axis in the $xy$-plane. For convenience, we define $\mu \equiv \cos \theta$. The slab consists of aligned grains with the alignment axis along the $x$-axis.

The unit vectors, $\hat{\theta}$ and $\hat{\phi}$, are used to denote the direction in increasing $\theta$ and $\phi$ respectively as shown in Fig.~\ref{fig:coordinates}. The Stokes parameter vector, $\vect{I} \equiv (I, Q, U, V)^{T}$, as viewed by an observer in the direction $\hat{n}$ are defined in the plane formed by $\hat{\theta}$ and $\hat{\phi}$, i.e., the sky or image plane. Following the definitions from \cite{Mishchenko2000}, positive Stokes $Q$ is polarization parallel to $\hat{\theta}$ and positive Stokes $U$ is polarization parallel to $\hat{\theta}-\hat{\phi}$. The linear polarization is
\begin{align}
    p_{l} &\equiv \dfrac{ \sqrt{Q^{2} + U^{2}} }{ I}
\end{align}
and the polarization angle is 
\begin{align}
    \zeta &\equiv \dfrac{1}{2} \text{atan2} \bigg(\dfrac{ - U}{ - Q} \bigg)
\end{align}
where the function atan2 computes the angle in the appropriate quadrant based on the signs of Stokes $Q$ and $U$ (see \citealt{Mishchenko2000}). As shown in Fig.~\ref{fig:coordinates}, the polarization angle $\zeta$ is measured from the direction of  $\hat{\phi}$ in the image plane and $\zeta$ increases in the clockwise direction as seen by the observer. Polarization parallel to $\hat{\phi}$ has $\zeta=0^{\circ}$ and polarization parallel to $\hat{\theta}$ has $\zeta=90^{\circ}$. Note that the convention in \cite{Mishchenko2000} is different from the IAU 1973 convention in which the position angle is defined from the North and increases in the counterclockwise direction. For Sections \ref{sec:plane_parallel} and \ref{sec:slab_results}, we use the \cite{Mishchenko2000} convention for consistency with the T-matrix code described below in Section~\ref{ssec:grain_model}.  

For the observer located at the direction ($\theta, \phi$), the radiation transfer equation for a ray is
\begin{align} \label{eq:fullrt}
    \mu \dfrac{d }{d z} \vect{I}_{\nu}(z, \theta, \phi) = &- \rho \matr{K}_{\nu}(\theta, \phi) \vect{I}_{\nu}(z, \theta, \phi)
        + \rho B_{\nu}(T) \vect{A}_{\nu}(\theta, \phi) \nonumber \\ 
        &+ \rho \oint \matr{Z}_{\nu}(\theta', \phi'; \theta, \phi) \vect{I}_{\nu}(z, \theta', \phi') d \Omega' 
        \text{,}
\end{align}
where $B_{\nu}$ is the black body radiation at some frequency $\nu$. $\matr{K}_{\nu}$, $\vect{A}_{\nu}$, and $\matr{Z}_{\nu}$ are the extinction matrix, absorption vector, and scattering matrix for the grain, and they depend on the grain orientation and material properties. For the remainder of the paper, we ignore the subscript $\nu$ for brevity, but note that all quantities related to intensity and opacity depend on frequency. $\matr{K}$ is a 4-by-4 matrix and $\vect{A}$ is a 4 element vector both of which depend on where the ray is headed ($\theta, \phi$). The scattering matrix $\matr{Z}$ is a 4-by-4 matrix which depends on where the scattered ray is headed ($\theta, \phi$) and also where the incoming ray was headed before scattering ($\theta', \phi'$). The opacity matrices depend on the dust model, which we will show in more detail below in Section \ref{ssec:grain_model}. The three terms on the right-hand-side of Eq.~(\ref{eq:fullrt}) are called the extinction, thermal emission, and scattering terms.

\begin{figure}
    \centering
    \includegraphics[width=0.7\columnwidth]{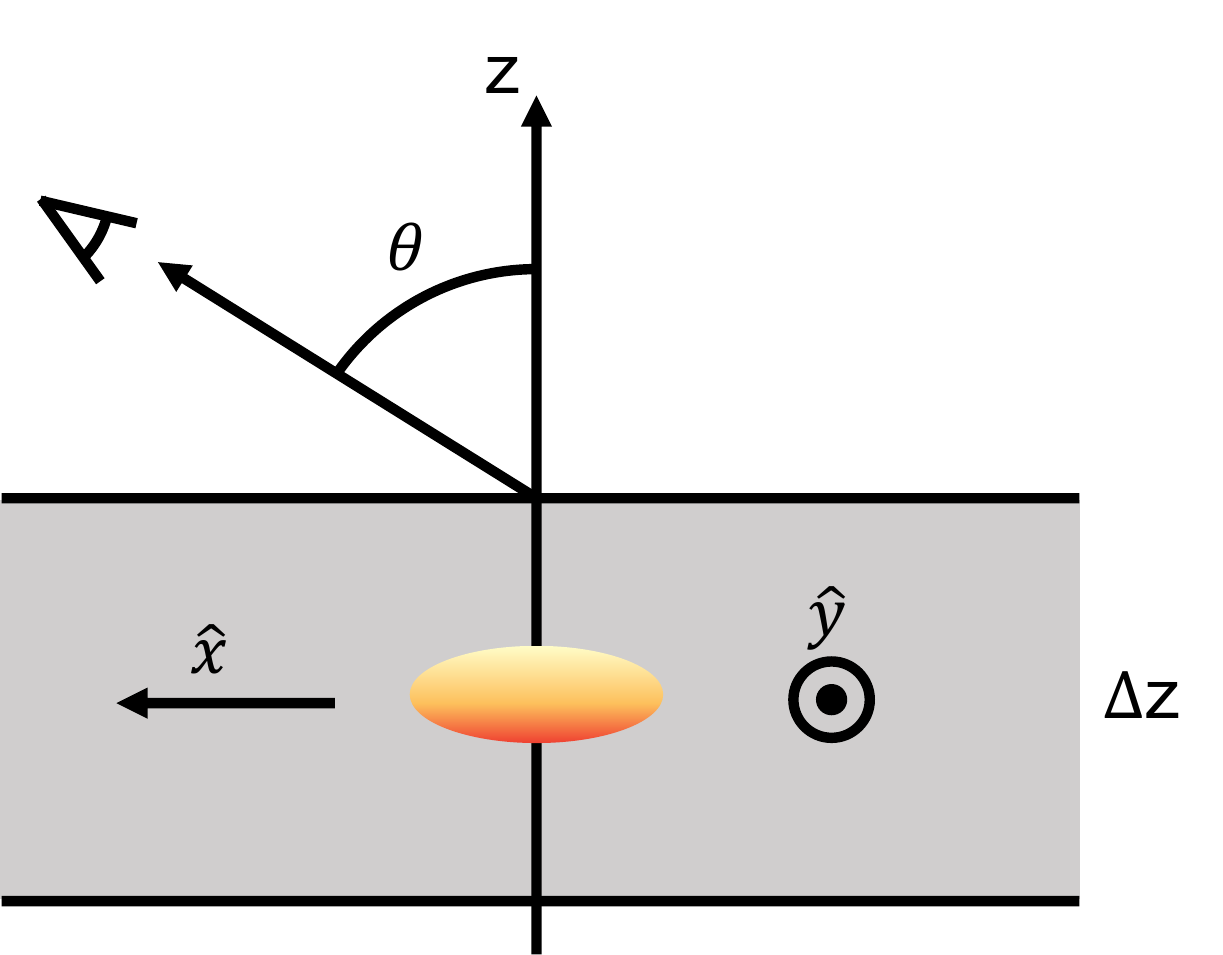}
    \caption{
        Schematic representation of the slab and the observer. In this example, the observer is located at some $\theta$ (between $0^{\circ}$ and $90^{\circ}$) and $\phi=0^{\circ}$. The thickness of the slab is $\Delta z$. The $x$-direction points to the left of the figure and the $y$-direction points out of the page. The slab is in gray which extends infinitely along the $x$- and $y$-direction, but finite in $z$. A cross section of a grain is represented by the colored ellipse and the shades of orange correspond to the colors of the prolate grain depicted in Fig.~\ref{fig:coordinates}. 
    }
    \label{fig:slabgeom}
\end{figure}

We are interested in solving the emergent intensity, which is the Stokes parameter vector that the observer sees from this slab (see Fig.~\ref{fig:slabgeom} for a schematic illustration). For an observer at $\mu>0$ (or $\theta \in [0^{\circ}, 90^{\circ})$), the emergent intensity is $\vect{I}(\Delta z,\theta,\phi)$. We assume that there is no impinging radiation from outside the slab, i.e., $\vect{I}(0,\theta,\phi)=0$ for $\mu>0$ and $\vect{I}(\Delta z,\theta,\phi)=0$ for $\mu<0$. For convenience, we define the Stokes $Q$ and $U$ normalized by Stokes $I$ respectively as
\begin{subequations} \label{eq:q_u}
    \begin{align}
        q &\equiv Q / I \\
        u &\equiv U / I \text{.}
    \end{align}
\end{subequations}

At some depth $z$, a grain scatters the incoming radiation $\vect{I}(z, \theta', \phi')$ to the observer at $(\theta, \phi)$ with the Stokes parameter $\matr{Z} \vect{I} d \Omega'$. Including scattering (the last term) in the radiative transfer equation greatly complicates the problem, which would otherwise be a simple integration. The amount of radiation towards the observer at each location no longer depends on just the local (non-radiation) quantities such as the temperature and density, but also the incoming radiation (the scattering term). However, knowing the incoming radiation means knowing the full radiation field. That includes the radiation towards the observer which is what we seek to solve in the first place. This is the ``chicken or egg'' effect that makes scattering problems difficult to solve. 

To solve the ``chicken or egg'' effect, we divide the radiation field into discrete grid points and iterate for a converging solution. This procedure is commonly called ``lambda iteration'' \citep{Mihalas1978}. We use the Gauss-Legendre quadrature to integrate the $\mu$ direction by dividing the $\mu$-grid into an even number of sampling points $N_{\mu}$. We consider a linear $\phi$-grid with $N_{\phi}$ points and a linear $z$-grid with $N_{z}$ points. Before each iteration, we use the T-matrix method to produce the opacities at the discrete angular points (details are described in the Section \ref{ssec:grain_model}). For all slab calculations below, we use $N_{\mu}=32$, $N_{\phi}=32$, and $N_{z}=128$. Higher number of sampling points did not change the results much which we demonstrate in Appendix~\ref{sec:convergence_gridpoints}. 

We begin the iteration by first assuming the radiation field is zero. In other words, we set $\vect{I}(z,\theta',\phi')=0$. In this case, the scattering term is zero, and we know the complete source term. We solve the differential equation numerically using the DELO-linear method \citep{Rees1989, Janett2017} to obtain a first approximation to the radiation field $\vect{I}(z,\theta,\phi)$. We then simply replace the incoming radiation in the scattering term with the approximated radiative field and solve the differential equation again for a new estimate of the radiation field. This procedure repeats for several iterations until the radiation field converges. The radiation field is converged if the maximum relative difference between the new and the prior iterations is less than $0.1\%$. For small grains with low albedo, it takes less than 10 iterations to converge. For large optical depth and grains with large albedo, it can take a few hundred iterations to converge. Appendix \ref{sec:consistency_check} shows an example of how the emergent intensity converges and the converged solution of Stokes $I$ is tested against the analytical solution assuming isotropic scattering. 

Once a converged solution to the radiation field is obtained, we can post-process to find the emergent intensity at any ($\theta$, $\phi$) which do not have to lie on the discrete grid. We use the dust model to evaluate the opacity matrices at the new ($\theta$, $\phi$) and use the discrete radiation field to evaluate the scattering term. A simple integration along the line of sight with the same boundary conditions will give us the emergent intensity.

\subsection{Grain Model} \label{ssec:grain_model}
The T-matrix method is a numerical modeling technique for light scattering by nonspherical particles of arbitrary size (see \citealt{Mishchenko1996} and \citealt{Mishchenko2000} for a review). We use the python package PyTMatrix\footnote{Leinonen, J., Python code for T-matrix scattering calculations. Available at \url{https://github.com/jleinonen/pytmatrix/}.} to calculate the amplitude matrix and scattering matrix for prolate grains \citep{Leinonen2014}\footnote{See \cite{Mishchenko2000} for details on calculating the extinction and absorption matrices from the amplitude and scattering matrices.}. As described in Section \ref{sec:introduction}, the choice of prolate grains is motivated by the work of \cite{Yang2019} and \cite{Mori2021} who demonstrated the potential for prolate grains to explain the azimuthal polarization pattern observed in HL Tau in Band 3. PyTMatrix is a python interface for a T-matrix FORTRAN code \citep{Mishchenko1994, Mishchenko1996_cylinder, Wielaard1997}\footnote{The FORTRAN code is available at \url{https://www.giss.nasa.gov/staff/mmishchenko/t_matrix.html}.}. 

The prolate grain geometry is characterized by its aspect ratio $s$ and size $a$. The ratio between the short axis to the axis of symmetry (i.e., the long axis) is defined as $s$. A prolate grain has $s <1$, while a spherical grain has $s = 1$. Its size is defined by the radius of the volume-equivalent sphere, i.e., the volume of the prolate grain is equal to the volume of a sphere with radius $a$.

We adopt the Disk Substructures at High Angular Resolution Project (DSHARP) mixture, which consists of 20\% water ice, 33 \% astronomical silicates, 7\% troilite, and 40\% refractory organics by mass \citep{Birnstiel2018}. The optical constants for water ice are from \cite{Warren2008}, astronomical silicates from \cite{Draine2003}, and troilite and refractory organics from \cite{Henning1996}. We assume the MRN size distribution of $n(a) \propto a^{-3.5}$ where $a$ is the grain size \citep{Mathis1977}. The distribution cuts off at a minimum grain size $a_{\text{min}}$ and a maximum grain size $a_{\text{max}}$. Since the results are not sensitive to $a_{\text{min}}$, we set $a_{\text{min}}=0.01 \mu$m \citep[e.g][]{Ricci2010, Kataoka2015}.

Since radiation transfer relies on the optical depth, it is convenient to define the average extinction opacity of the non-spherical grain for non-polarized light: 
\begin{align}
    \kext \equiv \dfrac{1}{4 \pi} \int K_{1,1}(\theta, \phi) d\Omega \text{ .}
\end{align}
We use this to define the vertical maximum optical depth of the slab as $\tau_{m} = \rho \kext \Delta z$.

\section{Plane-Parallel Slab Results} \label{sec:slab_results}
In this section, we discuss the results from the plane-parallel model looking at effects due to inclination, optical depth, and azimuthal variation to build intuition. We adopt $s=0.975$ consistent with the Band 3 polarization of $\sim 2\%$, i.e., the grains cannot be too elongated if the grains are well aligned. For illustration purposes, we pick set the observing wavelength, $\lambda$, at 1mm, representative of the ALMA wavelengths. The size parameter, $x \equiv 2 \pi a / \lambda$, of the maximum grain is set at $x = 0.4$ (or $a_{\text{max}}\sim 64 \mu$m). With the choice of $x$, the albedo of the adopted dust model is $\sim 0.4$ which makes the scattering polarization comparable to the thermal polarization and illustrates clearly the interplay between the scattering effects and the direct emission. The size parameter is small enough such that the grain is not too far from the dipole regime making it easier to understand as opposed to the Mie regime ($x=1$ or larger). In this paper, we will concentrate on the Stokes $Q$ and $U$ that describe the linear polarization. Stokes $V$ is also computed, but, since it is yet to be firmly detected, we will postpone a detailed exploration of this quantity to a future paper.  

\subsection{Inclination Effects} \label{ssec:inclination}
Inclination of a slab with only spherical grains induces polarization through scattering that is parallel the direction of $\hat{\theta}$ in Fig.~\ref{fig:coordinates} with positive Stokes $Q$ \citep{Yang2016_inc}. When considering aligned oblate grains in the single-scattering limit, \cite{Yang2016_oblate} showed that the thermal polarization superposes with the inclination-induced polarization due to scattering. We first look at how scattering of aligned grains produces polarization at different inclinations with the inclusion of multiple scattering. 

\begin{figure}
    \centering
    \includegraphics[width=\columnwidth]{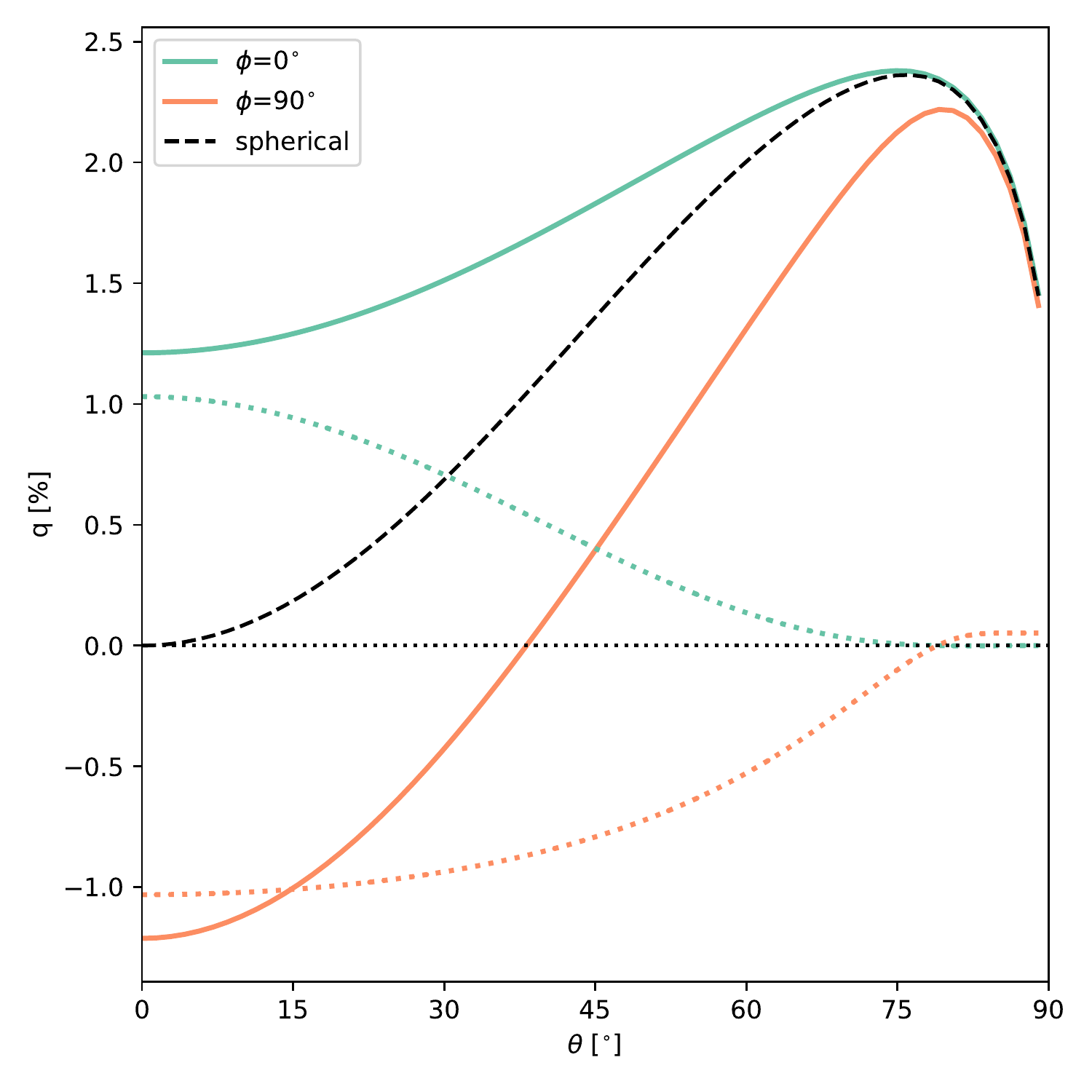}
    \caption{
        Linear polarization, expressed by $q \equiv Q/I$ (since $u\equiv U/I=0$), in percent as a function of inclination angle $\theta$ of the slab to the line of sight. The total vertical optical depth $\tau_{m}=1$, the maximum grain size parameter $x=0.4$, and the aspect ratio $s=0.975$. The green solid line is the emergent polarization at $\phi=0^{\circ}$ for scattering aligned grains, whereas the orange solid line is that at $\phi=90^{\circ}$. The colored dotted lines represent polarization from aligned grains without scattering. The black dashed line corresponds to the linear polarization for spherical grains. The horizontal dotted line is $q=0$. 
    }
    \label{fig:q_inc}
\end{figure}

We fix the vertical optical depth $\tau_{m}=1$ and show $q$ (the Stokes $Q$ normalized by Stokes $I$; Eq. (\ref{eq:q_u})) as a function of inclination for $\phi=0^{\circ}$ and $90^{\circ}$ in Fig.~\ref{fig:q_inc}. Results for other azimuthal angles are presented in Section \ref{ssec:azimuthal}. The prolate grains, as seen by the observer, are projected vertically for $\phi=0^{\circ}$ (see Fig.~\ref{fig:slabgeom} for an illustration) and horizontally for $90^{\circ}$. At these two azimuths, Stokes $U$ is zero because of the symmetry of the system. Thus, linear polarization is completely described by $q$. To help understand the polarization curve from scattering aligned grains, we also plot two limiting cases: the polarization for a slab of scattering spherical grains of the same effective size and the polarization for a slab of aligned grains without scattering.

The limiting case of polarization for spherical grains helps gauge the degree of the polarization induced by inclination itself. It is computed by setting $s=1.0$ while keeping the rest of the parameters the same. The polarization curve is near zero when $\theta$ is close to zero (face-on). When the inclination increases, $q$ is positive and increases up to some peak at around $\theta \sim 75^{\circ}$. The curve is similar to Fig.~2 of \cite{Yang2017} which considered a semi-infinite slab with single scattering. In Appendix~\ref{sec:consistency_check}, we illustrate with spherical grains how multiple scattering gradually adds Stokes $I$ and $q$ to the thermal emission until convergence.

For the non-scattering aligned grains ($s=0.975$), we calculate the emergent intensity through Eq.~(\ref{eq:fullrt}) without including the scattering term (though the scattering contribution to the extinction matrix remains). In other words, the grains are emitting polarized thermal radiation which undergoes dichroic extinction (photons are absorbed or scattered away), but none of the radiation is scattered into the line of sight of the observer. At $\phi=0^{\circ}$, $q$ of the non-scattering aligned grains is positive, because the long axis of the inclined grain looks vertical to the observer which produces positive Stokes $Q$. As the inclination increases, the observer views the prolate grain more pole-on (illustrated in Fig.~\ref{fig:slabgeom}), with a reduced degree of elongation which leads to less polarization. Additionally, dichroic extinction along the line of sight increases due to an increase in path length, which decreases the polarization even further. At $\phi=90^{\circ}$, $q$ is negative, because the long axis of the prolate grain looks horizontal (parallel to $\hat{\phi}$ in Fig.~\ref{fig:coordinates}) to the observer which produces negative Stokes $Q$. As the inclination increases, the polarization just from the shape of the grain does not change. Instead, the decrease in $q$ is entirely due to the increase in path length along the sight line and thus the dichroic extinction. For both viewing azimuths, the magnitude of polarization from direct thermal emission decreases with increasing inclination which is opposite from scattering polarization of spherical grains (at least when $\theta<75^{\circ}$). 

For aligned grains including scattering, the polarization remains completely positive at any inclination for $\phi=0^{\circ}$. When viewed at $\phi=90^{\circ}$, the polarization changes direction from being perpendicular to $\hat{\theta}$ (negative $q$) to being parallel to $\hat{\theta}$ (positive $q$). We can understand the inclination effects for scattering of aligned grains from the limiting cases of scattering from spherical grains and of non-scattering aligned prolate grains. 

At low inclination, polarization from scattering aligned grains is similar to the polarization from non-scattering aligned grains (essentially thermal polarization). At the same time, polarization from scattering spherical grains reaches $0$, which means scattering is inefficient at producing the inclination-induced polarization. However, there is a slight difference between scattering aligned grains and non-scattering aligned grains at $i=0^{\circ}$, which arises from scattering due to the differential cross section of the grain instead of inclination explained by the following. In the case of spherical grains, radiation coming from the $x$- and $y$-axis in Fig.~\ref{fig:coordinates} scatter by $90^{\circ}$ to the observer at $i=0^{\circ}$ (or along the $z$-axis), which makes the radiation maximally polarized with a direction perpendicular to the respective scattering plane. Since the scattering cross section is the same for both, the two polarization cancel out, and the resulting polarization that reaches the observer is 0. The scenario is different for the prolate grain. The radiation coming along the pole of the grain (along the $x$-axis) sees a smaller scattering cross section than the cross section seen by radiation traveling perpendicular to the grain long axis (along the $y$-axis). Even though the scattered radiation is maximally polarized for incoming photons from both directions, a larger fraction of the photons coming along the $y$-axis is scattered than that along the $x$-axis. The resulting scattered polarization has the same orientation as that of the thermal polarization of the grain, which serves as a boost to the polarization (see \citealt{Yang2016_oblate} for a description of the oblate case).

At higher inclinations, the thermal polarization decreases due to increasing dichroic extinction, while the polarization of scattering aligned grains is similar to the polarization with only spherical grains at both $\phi=0^{\circ}$ and $90^{\circ}$. This means polarization by scattering becomes dominant when the inclination is large enough. For $\phi=90^{\circ}$, the change in the sign of $q$ (and consequently the $90^{\circ}$ change in the polarization direction) is due to the changing balance between the two polarization mechanisms as the inclination angle and the associated optical depth along the sight line changes.

\subsection{Optical Depth Effects} \label{ssec:optical_depth}
\begin{figure*}
    \centering
    \includegraphics[width=\textwidth]{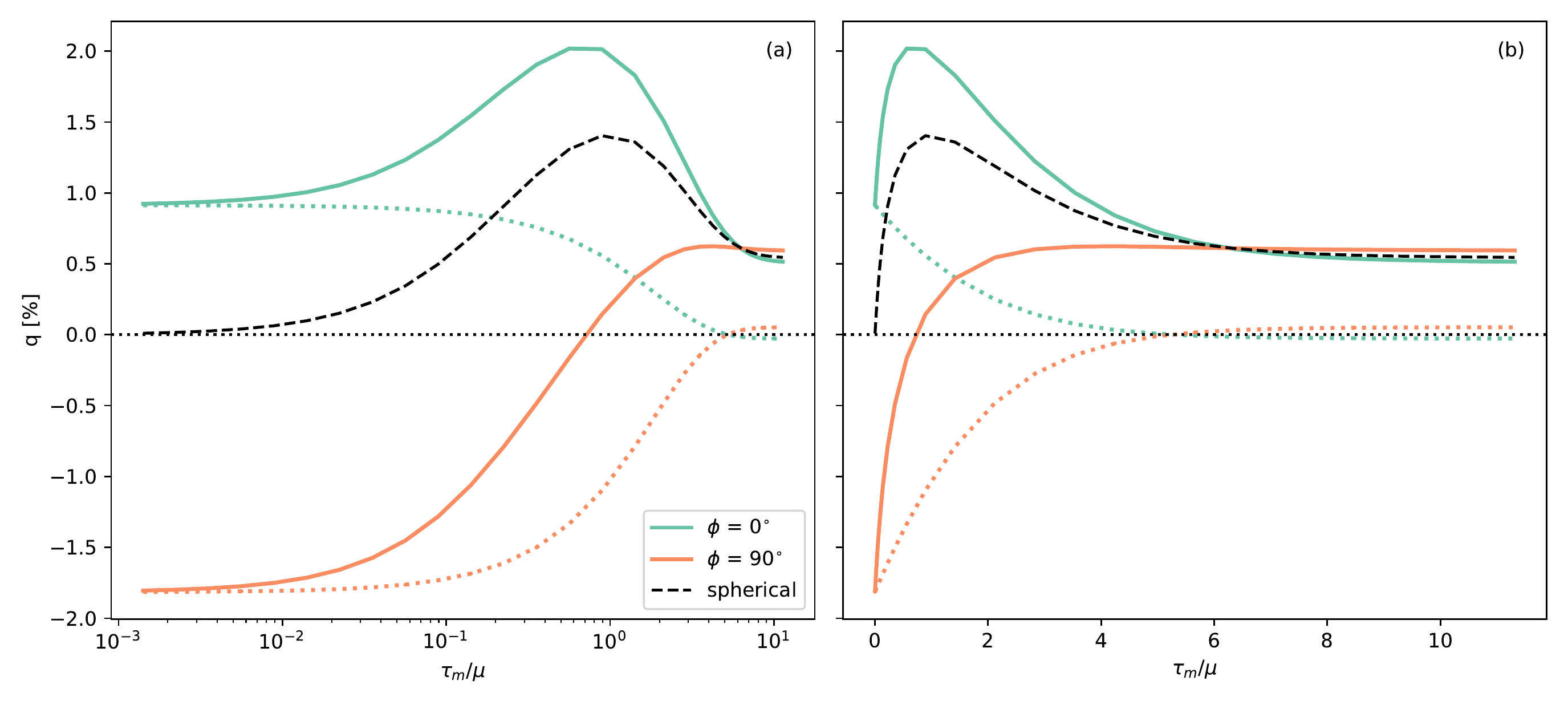}
    \caption{
        Linear polarization defined by $q \equiv Q/I$ in percent as a function of total optical depth along the line of sight $\tau_{m}/\mu$. Panel (a) is the same as panel (b) except the horizontal axis is in log scale on the left and in linear scale on the right to easily view the optically thin and optically thick regimes. The solid green line is the emergent polarization at $\phi=0^{\circ}$ for scattering aligned grains, whereas the orange line is that at $\phi=90^{\circ}$. The colored dotted lines represent polarization from aligned grains without scattering. The black dashed line corresponds to the linear polarization for spherical grains. The horizontal dotted line is $q=0$.
    }
    \label{fig:q_opt}
\end{figure*}

To isolate the optical depth effects, we will fix the inclination angle to $\theta=45^\circ$ and vary the total (vertical) optical depth of the slab $\tau_m$ in this section. The variation in optical depth can come from either the spatial (i.e., radial) variation of the dust distribution at a single wavelength or a change of observing wavelength at the same disk location. The results are illustrated in Fig.~\ref{fig:q_opt}, where we plot the linear polarization fraction $q$ as a function of the total optical depth of the slab along the line of sight $\tau_m/\mu$ (where $\mu=\sqrt{2}/2$ for $\theta=45^\circ$) at $\phi=0^\circ$ and $90^\circ$. Similar to Section \ref{ssec:inclination}, we also show $q$ from non-scattering aligned grains and from scattering spherical grains. To better view the optically thin and optically thick limits, we show the same polarization results, but plot the optical depth in linear and log scales. 

For the case of non-scattering aligned grains at both $\phi=0^{\circ}$ and $90^{\circ}$, the polarization is non-zero at low optical depth and determined simply by its projected shape \citep{Draine1984}. As the optical depth increases, dichroic extinction reduces the polarization from direct thermal emission monotonically \citep{Hildebrand2000}.  

For scattering spherical grains, we see that at low $\tau_{m}/\mu$, the polarization approaches zero because the photons hardly scatter. The polarization peaks at an optical depth of $\sim 1$ and asymptotes to a constant value at large optical depths. This result is qualitatively similar to Fig. 3 in \cite{Yang2017} which considered only single-scattering. 

The polarization of scattering aligned grains is a mix of these two limiting cases. At low optical depth, polarization from scattering aligned grains follows the thermal polarization because there is a lack of scattering for inclination effects to produce significant polarization. As the optical depth increases, scattering of photons is more likely to occur and produces polarization induced by inclination. At the same time, thermal radiation undergoes more dichroic extinction as it travels through the optically thick slab and thermal polarization decreases. As a result, inclination-induced polarization dominates at large optical depth. For $\phi=0^{\circ}$, polarization stays positive entirely, because the thermal polarization and the inclination-induced polarization both produce positive Stokes $Q$. For $\phi=90^{\circ}$, polarization is initially negative at low $\tau_{m}/\mu$, but it becomes positive at higher optical depth. There is a cross-over point (or polarization reversal) at $\tau_{m} / \mu \sim 0.7$. It makes physical sense because the thermal polarization (negative $q$) is opposite to the inclination-induced polarization (positive $q$) and the former dominates at low optical depth, while the latter dominates at high optical depth. The cross-over point is not limited to $\tau_{m}/\mu$ of order unit, but instead depends on the aspect ratio or the grain. For example, a spherical grain cannot directly emit any $q$ and the cross-over happens effectively at $\tau_{m}/\mu=0$, while a slab of highly elongated prolate grains will require much larger optical depth to take out the large intrinsic polarization. 

% also explain how to boost polarization due to scattering other than just mention that inclination produces this effect

% fix optical depth=10, show polarization for scattering aligned grains and vary aspect ratio 
At high optical depths, there exists a slight deviation in polarization between the spherical grains and that of scattering aligned grains. The polarization at $\phi=0^{\circ}$ is slightly less than the polarization of the spherical grains slab, while the polarization at $\phi=90^{\circ}$ is slightly higher. The deviation is on the order of the deviation of polarization of non-scattering aligned grains from zero. As we demonstrate below, this is a result of the incomplete cancellation between the polarization produced by dichroic extinction and that by thermal emission. 

Consider the non-scattering aligned grains case of an observer along $\phi=0^{\circ}$ or $90^{\circ}$. Since the grain does not emit Stokes $U$ or $V$ in this frame (i.e., $A_{3}=A_{4}=0$), we only have to solve for Stokes $I$ and $Q$. The thermal polarization from a grain is simply $A_{2}/A_{1}$. Also, in this frame, only two elements of the extinction matrix $\matr{K}$ are independent which we define: $K_{1}\equiv K_{11}=K_{22}$ and $K_{2}\equiv K_{12}=K_{21}$. The resulting radiation transfer equation along path $s$ becomes
\begin{align}
    \dfrac{d }{ d s} \begin{bmatrix}
        I \\
        Q
    \end{bmatrix} 
    = - \rho 
        \begin{bmatrix}
            K_{1} & K_{2} \\
            K_{2} & K_{1}
        \end{bmatrix} 
        \begin{bmatrix}
            I \\
            Q
        \end{bmatrix}
        + \rho B
        \begin{bmatrix}
            A_{1} \\
            A_{2}
        \end{bmatrix}
\end{align}
which is a set of first order nonhomogeneous differential equations. 

Since scattering is not involved, we can easily obtain an analytical solution for $[I,Q]^{T}$. The eigenvalues for the extinction matrix are $K_{\pm} \equiv K_{1} \pm K_{2}$. The complete solution assuming no external radiation as a boundary condition is
\begin{align}
    \begin{bmatrix}
        I(s) \\
        Q(s)
    \end{bmatrix}
    =
    &- \dfrac{B}{2}
    \begin{bmatrix}
        e^{-\rho K_{+}s} & e^{-\rho K_{-}s} \\
        e^{-\rho K_{+}s} & - e^{-\rho K_{-}s} 
    \end{bmatrix}
    \begin{bmatrix}
        A_{+} / K_{+}\\
        A_{-} / K_{-}
    \end{bmatrix}
    \nonumber \\
    &+ 
    \dfrac{B}{ K_{+} K_{-} }
    \begin{bmatrix}
        K_{1} & - K_{2} \\
        - K_{2} & K_{1}
    \end{bmatrix}
    \begin{bmatrix}
        A_{1} \\
        A_{2}
    \end{bmatrix}
\end{align}
where $A_{\pm} \equiv A_{1} \pm A_{2}$. In the optically thin limit, we retrieve $Q/I = A_{2} / A_{1}$ as expected from a single grain. In the optically thick limit, the polarization is
\begin{align} \label{eq:q_therm_opt_thick}
    \dfrac{Q}{I} = \dfrac{ \dfrac{A_{2}}{A_{1}} - \dfrac{ K_{2} }{ K_{1} } }{ 1 - \dfrac{ K_{2} }{ K_{1} } \dfrac{ A_{2} }{ A_{1} } } \sim \dfrac{A_{2}}{A_{1}} - \dfrac{K_{2}}{K_{1}} 
    %= \dfrac{  K_{1} A_{2} - K_{2} A_{1} + }{ K_{1} A_{1} - K_{2} A_{2} } \text{ .}
\end{align}
where the approximation on the right-hand-side applies because $A_{2}/A_{1}$ is only a few percent for grains with small aspect ratio and the two quantities, $A_{2}/A_{1}$ and $K_{2}/K_{1}$, are comparable. Note that $K_{2}=0$ and $A_{2}=0$ for spherical grains. Eq.~(\ref{eq:q_therm_opt_thick}) simply means the prolate grains emit thermal polarization, $A_{2}/A_{1}$, but the same grains attenuate the polarization through their dichroic extinction, $K_{2}/K_{1}$. In the limit of small grains when the scattering opacity is negligible, $A_{i}$ are equal $K_{i}$, which makes the polarization in the optically thick limit zero (e.g., \citealt{Hildebrand2000}). However, if the dichroic extinction does not perfectly attenuate the thermal polarization, then we get residual polarization even in the optically thick limit\footnote{Even if $A_{i}=K_{i}$, residual polarization can also be achieved when there is a temperature gradient \citep{Yang2017, Lin2020_poldt}}. For the chosen grains of $x=0.4$, the dichroic extinction takes out more polarization than the grains themselves can emit which leaves the polarization perpendicular to its thermal polarization. Note that the polarization in this limit, as expressed by Eq.~(\ref{eq:q_therm_opt_thick}), is typically small.

% Note that the polarization reversal shown here is different from the temperature gradient effects demonstrated in \cite{Lin2020_poldt}. 

\subsection{Azimuthal Variation} \label{ssec:azimuthal}
\begin{figure*}
    \centering
    \includegraphics[width=\textwidth]{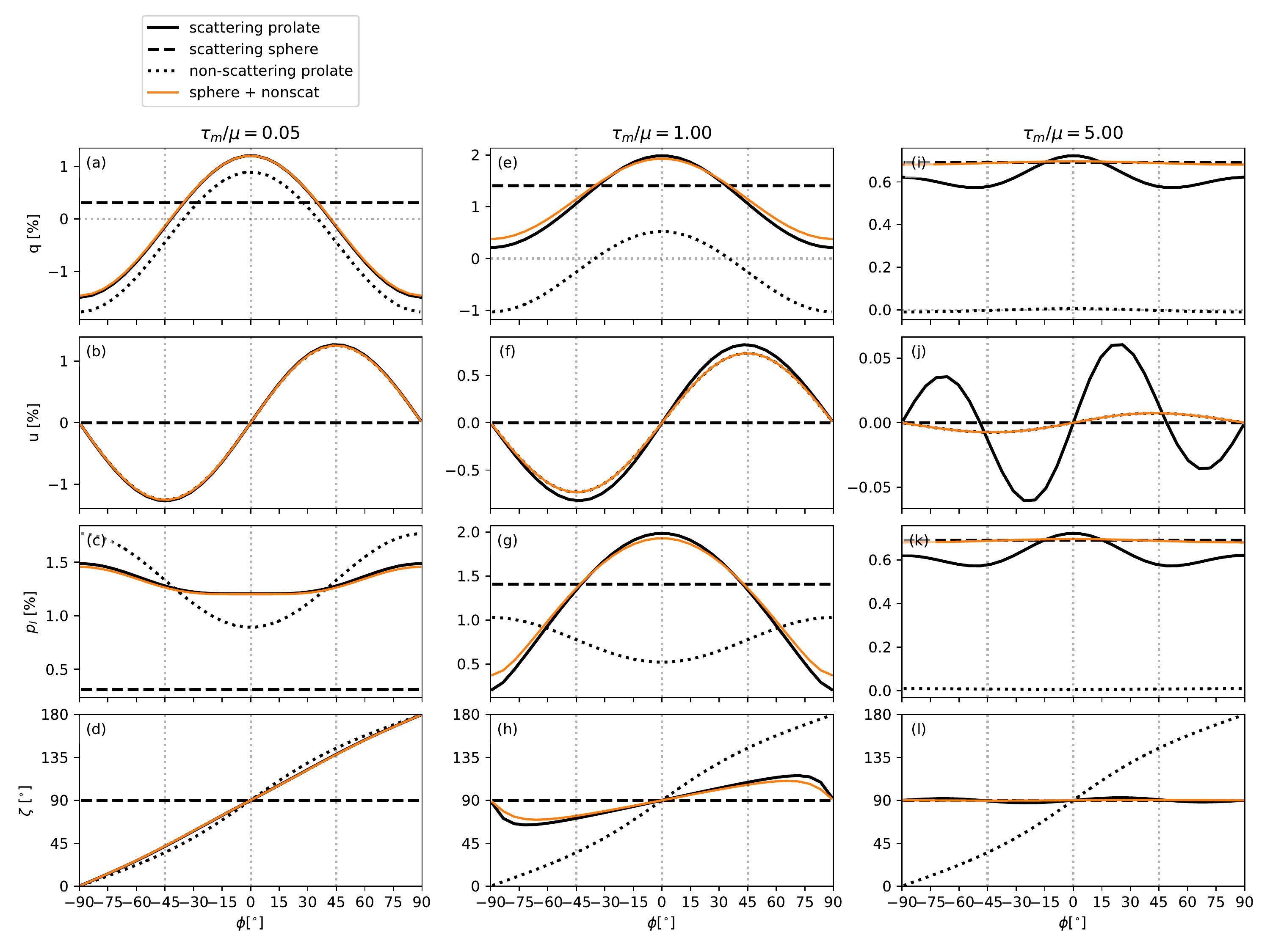}
    \caption{
        The azimuthal profiles for various polarization properties at $\theta=45^{\circ}$ with different total optical depth. From the top to bottom, the polarization properties are $q$ (the normalized Stokes $Q$), $u$ (the normalized Stokes $U$), linear polarization fraction all in percent, and polarization angle in degrees. Each column corresponds to a different total optical depth $\tau_{m}/\mu$ along the line of sight. The black solid lines are the results from the full calculation of scattering of aligned prolate grains. The dashed lines are scattering spherical grains. The dotted lines correspond to aligned grains without scattering. The orange solid lines are linear additions of $q$ from non-scattering aligned grains and $q$ from scattering spherical grains (likewise for $u$). The grey dotted lines are guidelines. The grey horizontal dotted lines for the first row is where $q=0$. The grey vertical dotted lines mark $\phi=-45^{\circ}$, $0^{\circ}$, and $45^{\circ}$. Note the change of the vertical range for $q$ and $u$ (the first and second rows) and the polarization fraction (the third row) for different optical depths (i.e., different columns).
    }
    \label{fig:azi_prof_taum}
\end{figure*}

Understanding the azimuthal variation of polarization requires us to consider $\phi$ other than the two specific azimuths discussed thus far. We use the same slab above, but fix the inclination at $45^{\circ}$. Since the system is no longer mirror symmetric in general, we need both Stokes $Q$ and $U$ to describe the linear polarization instead of just Stokes $Q$. Each row in Fig.~\ref{fig:azi_prof_taum} shows how the $q$ and $u$ vary as a function of azimuth along with $p_{l}$ and $\zeta$. Each column corresponds to a different $\tau_{m}/\mu$ to see how the azimuthal profile changes with different optical depth. We plot only from $\phi=-90^{\circ}$ to $90^{\circ}$ since the system is symmetric under $180^{\circ}$ rotation for our setup. 

We also plot the case of scattering spherical grains and non-scattering aligned grains. For scattering spherical grains, the emergent intensity does not depend on $\phi$. Thus, all Stokes parameters are constant across $\phi$. $q$ (and consequently Stokes $Q$) is positive due to inclination-induced polarization. On the other hand, $u$ is entirely 0, because the spherical grain cannot emit any polarized emission directly and scattering cannot produce Stokes $U$ due to of the symmetry of the system. As such, the polarization angle is always $90^{\circ}$. 

It is helpful to understand the non-scattering aligned grains starting with the most optically thin case $\tau_{m}/\mu=0.05$. As seen from the $\phi=0^{\circ}$ and $90^{\circ}$ shown in Sections \ref{ssec:inclination} and \ref{ssec:optical_depth}, the sign of $q$ depends on the orientation of the grain. Shown in Fig.~\ref{fig:azi_prof_taum}, the switch in the sign occurs roughly at $\phi \sim 35^{\circ}$ as the projected grain looks slightly more horizontal than vertical. Also at $\phi\sim45^{\circ}$, $u$ is maximized since the projected long axis of the grain looks diagonal with $\zeta \sim 135^{\circ}$ (Fig. \ref{fig:azi_prof_taum}b). The minimum $u$ happens at $\phi \sim - 45^{\circ}$ when the grain also looks diagonal but with $\zeta\sim 45^{\circ}$.\footnote{Considering a dipole in the optically thin limit, the switch of the sign of $q$ is at $\phi=\pm35^{\circ}$ when $\theta=45^{\circ}$. The maximum and minimum for $u$ is at $\phi=45^{\circ}$ and $-45^{\circ}$ respectively regardless of $\theta$.} At other $\tau_{m}$, the azimuthal variations of $q$ and $u$ from the thermal emission by non-scattering prolate grains are similar except that the magnitudes of $q$ and $u$ are decreased because of an increased dichroic extinction at a higher optical depth. 

For scattering aligned grains, in the optically thin case of $\tau_{m} / \mu = 0.05$, the $q$ and $u$ (Fig.~\ref{fig:azi_prof_taum}a and b) are similar to those of the non-scattering aligned grains except there is a positive offset of $q$ relative to the non-scattering aligned grain. The amount of offset is similar to $q$ of scattering spherical grains. The $u$ in this case is completed dominated by the direct emission from the aligned prolate grains with no contribution from scattering. In Fig.~\ref{fig:azi_prof_taum}c, including scattering to aligned grains decreases the degree of azimuthal variation of $p_{l}$. This is due to the addition and canceling effects between inclination-induced polarization from scattering and thermal polarization at different azimuth. The inclination adds to the positive $q$ thermal polarization at $\phi=0^{\circ}$ and decreases $q$ at $\phi=90^{\circ}$ as shown in Section~\ref{ssec:inclination}. At $\phi = \pm 45^{\circ}$, $u$ largely determines $p_{l}$ meaning that adding scattering does not alter $p_{l}$ since $u$ from scattering is 0. The slight contribution from $q$ due to scattering creates the deviation of polarization angle $\zeta$ from the non-scattering aligned grains case (Fig.~\ref{fig:azi_prof_taum}d). Specifically, the polarization angle deviates towards $90^{\circ}$ (or parallel to $\hat{\theta}$; Fig.~\ref{fig:coordinates}). 

For the unity optical depth case of $\tau_{m}/\mu = 1$, $q$ increased relative to $\tau_{m}/\mu=0.05$ as expected since scattering becomes more prominent and contributes positive $q$ due to inclination (Fig. \ref{fig:azi_prof_taum}e). Due to this increase in $q$, $p_{l}$ now peaks at $\phi=0^{\circ}$ and reaches a minimum at $\phi=\pm 90^{\circ}$ (Fig. \ref{fig:azi_prof_taum}g). The location of the peak and trough is opposite to the $\tau_{m}/\mu=0.05$ case (Fig. \ref{fig:azi_prof_taum}c). It is likely that there exists an optical depth $\tau_m/\mu$ between 0.05 and 1 that makes the polarization fraction nearly independent of the azimuthal angle $\phi$. Also, $p_{l}$ can become roughly flat across azimuth when the level of scattering is just enough at some optical depth. The polarization angle $\zeta$ shown in Fig.~\ref{fig:azi_prof_taum}h evolves towards $90^{\circ}$ as the optical depth increases. 

When the optical depth reaches $\tau_{m}/\mu=5$, $q$ of the scattering aligned grains varies around the level of $q$ of scattering spherical grains in Fig. \ref{fig:azi_prof_taum}i. In contrast, $q$ from non-scattering aligned grains is low. Thus, we see that the inclination-induced polarization due to scattering dominates the azimuthal profile. The thermal polarization provides a deviation to $q$, but there is an additional deviation particularly at $\phi\sim 45^{\circ}$. In Fig. \ref{fig:azi_prof_taum}j, $u$ from scattering aligned grains does not resemble that from non-scattering aligned grains as was the case for lower optical depths (Fig.~\ref{fig:azi_prof_taum}b,f). Since $u$ is an order of magnitude smaller than $q$, the effect on $p_{l}$ is small. As a result, $p_{l}$ mainly follows $q$ and the polarization angle is mainly constant at $90^{\circ}$ (Fig.~\ref{fig:azi_prof_taum}k,l). 

From Fig.~\ref{fig:q_inc}, \ref{fig:q_opt}, and \ref{fig:azi_prof_taum}, we can observe that $q$ of scattering spherical grains and $q$ of non-scattering aligned grains roughly add together linearly to produce $q$ of scattering aligned grains. For direct comparison, we add the $q$ and $u$ quantities of the non-scattering aligned grains with those of scattering spherical grains in Fig.~\ref{fig:azi_prof_taum} and find close agreement with the proper scattering aligned grains case for $\tau_{m}/\mu=0.05$ and $1$. The linearity is because the scattering term and thermal emission terms add linearly as the source term. Furthermore, the prolate grain of $s=0.975$ is nearly spherical which makes the scattering matrix not too different from the spherical case \citep[e.g.][]{Kirchschlager2020} and that is why inclination produces similar polarization behavior for both. Scattering of a non-spherical grain can cause deviations (for example, at $i=0^{\circ}$ in Fig.~\ref{fig:q_inc}) though that is small compared to what inclination can produce (however, see Appendix~\ref{sec:elongated_prolate} for a case with highly elongated prolate grains). 

The linearity breaks down when the optical depth is large (Fig.~\ref{fig:azi_prof_taum}i): scattering changes the radiation field in the slab drastically and has a strong impact on the scattering source term. This impact is non-local and is not important at low optical depth, but becomes obvious at high optical depths. Thus, to a good approximation, particularly in the optically thin and moderately optically thick regimes, the bulk of the polarization is the superposition of the thermal polarization from the non-spherical grain and the inclination-induced polarization approximated by spherical grains (see Appendix~\ref{sec:larger_prolate_azimuthal} for a case with larger size parameter).

%% maybe a section on grain size 

\section{Application to HL Tau} \label{sec:hl_tau}
With the physical intuition gained from the plane-parallel model, we turn to actual observations. The dust disk of HL Tau is found to be highly settled \citep{Pinte2016} meaning that the vertical extent is much smaller than the extent of the horizontal distribution of material. The temperature should be roughly vertically isothermal where the bulk of the dust is located. Thus, we can approximate each patch of the disk locally with a plane-parallel model. By assuming the disk is axisymmetric and geometrically thin, we can piece together 2D images of the disk in the plane of sky in Section~\ref{ssec:trend_2d_model_QU} and understand the polarization transition across radius for at a given image and across multiple wavelengths by varying the optical depth. Following Section \ref{sec:slab_results}, the modeling effort cumulates in a simple empirical method that can be used to disentangle the contributions to the observed disk polarization from scattering and thermal emission (see Section \ref{ssec:empirical_pol_azi} below). 

\subsection{Observations} \label{ssec:observations}
\begin{figure*}
    \centering
    \includegraphics[width=\textwidth]{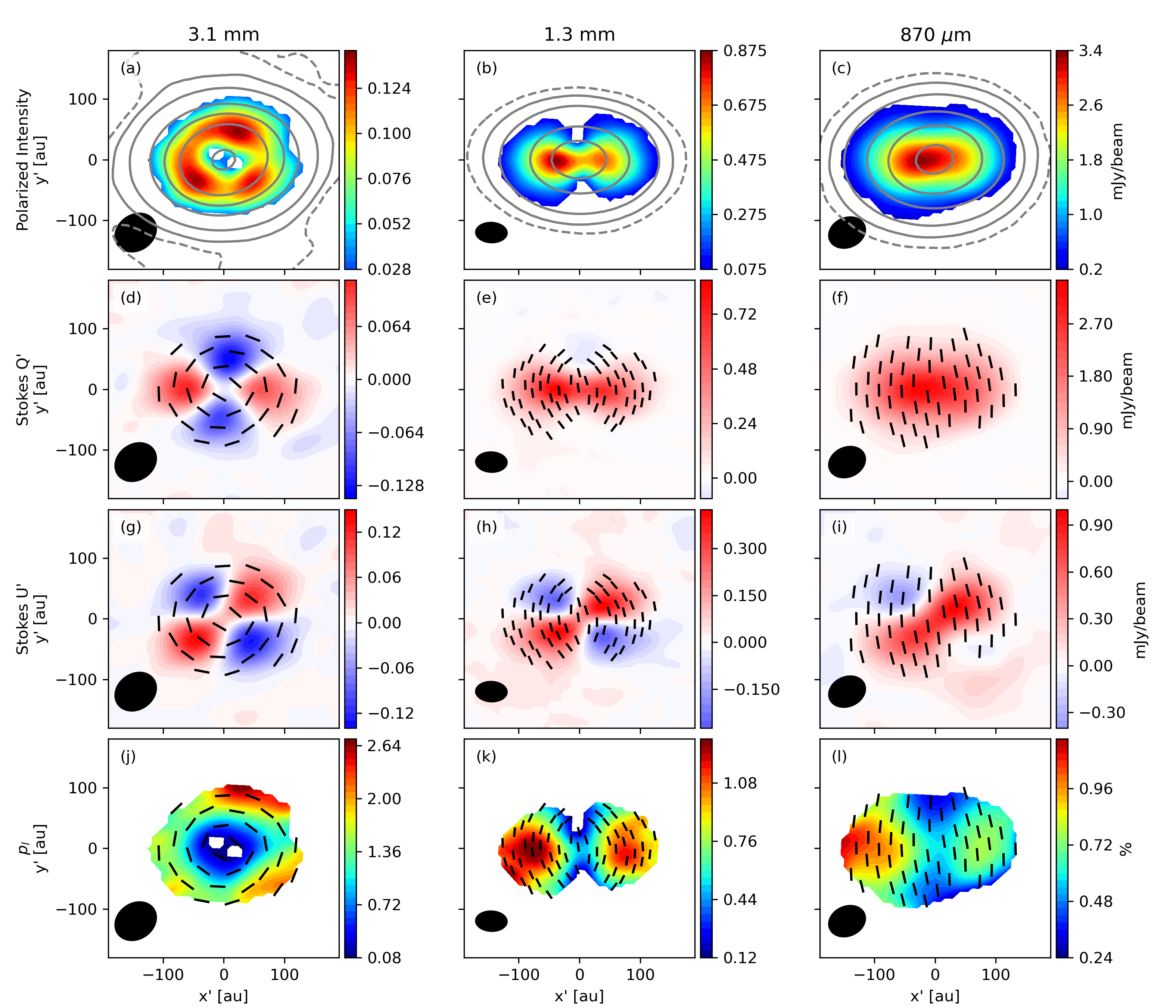}
    \caption{
        The polarization images rotated to the principle frame. The columns from the left to right are images from Band 3, 6, and 7. The rows from top to bottom are the linear polarized intensity, Stokes $Q'$, Stokes $U'$, and $p_{l}$. The first three are in mJy beam$^{-1}$, while $p_{l}$ is in percent. The contours in the first row are Stokes $I'$ in levels of [3, 15, 60, 240, 500]$\times \sigma_{I}$, and the dashed contour is the first level. Polarized intensity below a signal-to-noise of 3 are not plotted. The black ellipses in the lower left are the beam sizes. The vectors from the second to the last row represent the polarization angle, and the length of the vectors are uniform. 
    }
    \label{fig:obs_rot}
\end{figure*}

% but convolve them to the smallest common circular beam of $0.51\arcsec$ to avoid any systematic differences between images due to resolution effects.
We use the HL Tau multiwavelength images shown in \cite{Stephens2017}, but we rotate the image such that the disk minor axis is along the vertical direction of the image (Fig.~\ref{fig:obs_rot}). For convenience, we will call this the ``principle frame'' because the major and minor axes form the horizontal and vertical axes of the image for an inclined axisymmetric flat disk. The disk inclination $i$ is $46.7^{\circ}$ ($i=0$ means face-on), and the position angle of the disk minor axis $\eta$ is $48.2^{\circ}$ East-of-North \citep{almapartnership2015}. The spatial coordinates of the image in the principle frame $(x', y')$ is related to the coordinates of the original image by 
\begin{align}
    \begin{bmatrix}
        x' \\
        y'
    \end{bmatrix}
    = 
    \begin{bmatrix}
        \cos \eta & \sin \eta \\
        - \sin \eta & \cos \eta 
    \end{bmatrix}
    \begin{bmatrix}
        - x \\
        y \\
    \end{bmatrix}
\end{align}
where $x$ and $y$ are the RA and DEC relative to the disk center of the original image. We adopt the distance of 147 pc \citep{Galli2018}. 

The coordinate rotation changes the reference frame for the Stokes parameters as well. We use $\vect{I}$ to represent the original Stokes parameters as observed by ALMA and use primes, $\vect{I}'$, to denote the Stokes parameters in the principle frame. The Stokes parameters are related by 
\begin{align}
    \begin{bmatrix}
        Q' \\
        U'
    \end{bmatrix}
    = 
    \begin{bmatrix}
        \cos 2\eta & \sin 2 \eta \\
        - \sin 2 \eta & \cos 2 \eta 
    \end{bmatrix}
    \begin{bmatrix}
        Q \\
        U
    \end{bmatrix}
\end{align}
while Stokes $I$ equal Stokes $I'$. Since inclination-induced polarization of a disk produces polarization parallel to the disk minor axis, such polarization will exhibit positive Stokes $Q'$. The deviation of the polarization angle away from the disk minor axis will show up as Stokes $U'$, which makes it easy to identify in the rotated frame. Fig.~\ref{fig:obs_rot} shows the resulting linear polarized intensity, Stokes $Q'$, and Stokes $U'$ all in mJy beam$^{-1}$ (note this is not $q'$ and $u'$) and the polarization fraction image with the polarization vectors to denote the polarization angle.

Assuming the dust disk is geometrically thin \citep[e.g.][]{Pinte2016}, we can deproject the disk with a known inclination $i$ and obtain the azimuthal profiles by relating the spatial coordinate to the radius and azimuth:
\begin{align}
    x' &= r \cos \phi_{d} \\
    y' &= r \sin \phi_{d} \cos i
\end{align}
where $r$ is the radius. $\phi_{d}$ is the azimuth of the disk with $\phi_{d}=0^{\circ}$ starting along the right major axis going counterclockwise.

\subsection{Qualitative Trend Using a 2D Model} \label{ssec:trend_2d_model_QU}
Before fitting the data quantitatively, we will benefit from visualizing the slab models in the form of an image and discuss the main features. Since the disk is geometrically thin, we can approximate each point in the disk image as an independent slab, and we piece together multiple slab calculations to form the image. 

As described in Section \ref{sec:introduction}, we assume the prolate grains are aligned azimuthally around the disk. As such, we let the $x$-axis of the slab follow the direction opposite to increasing $\phi_{d}$ and let the $y$-axis of the slab point away from the disk center. The azimuth of the slab is related to the azimuth of the disk by 
\begin{align} \label{eq:phi_to_phi_d}
    \phi = 2\pi - \phi_{d} \text{ .}
\end{align}
The inclination of the disk corresponds to the polar angle of the slab $i=\theta$. Due to the different definitions of the Stokes parameters, Stokes $U$ (as defined by ALMA according to the IAU 1973 convention) differs from the Stokes $U$ (as defined by \citealt{Mishchenko2000}) of the plane-parallel slab by a negative sign. For direct comparisons, we convert the Stokes parameters of the slab \citep{Mishchenko1994} to the Stokes parameters of ALMA. 

%For clarity, we define $\zeta'$ as the polarization angle defined by Stokes $U$'.\footnote{Polarization parallel to the disk minor axis will have $\zeta'=0^{\circ}$. This is $90^{\circ}$ different from the slab definition, but to avoid confusion we will not  }

For multiwavelength polarization images like that of HL Tau, the optical depth decreases at longer wavelength, because the opacity of the grains decreases. At the same time though, the albedo varies and the refractive index also has a wavelength dependence. To isolate the effects just from optical depth, we will consider an image at $\lambda = 1$mm with grains of a maximum size parameter of $x=0.4$, which keeps the albedo and refractive index fixed, but we scale the surface density up and down to change the optical depth. We adopt a fairly simplistic but representative surface density profile \citep{Lyndenbell1974} and parameterize the dust surface density by 
\begin{align}
    \Sigma(r) = \dfrac{ \tau_{0} \cos{i} }{ \kext } \bigg( \dfrac{r}{ 100 \text{au} } \bigg)^{-0.5} \exp{ \bigg[ - \bigg( \dfrac{r}{100 \text{au}} \bigg)^{1.5} \bigg] }
\end{align}
where $r$ is the radius in au and $\tau_{0}$ is a characteristic optical depth. The temperature profile is 20 K at 100 au and goes as $r^{-0.5}$. 

\begin{figure*}
    \centering
    \includegraphics[width=\textwidth]{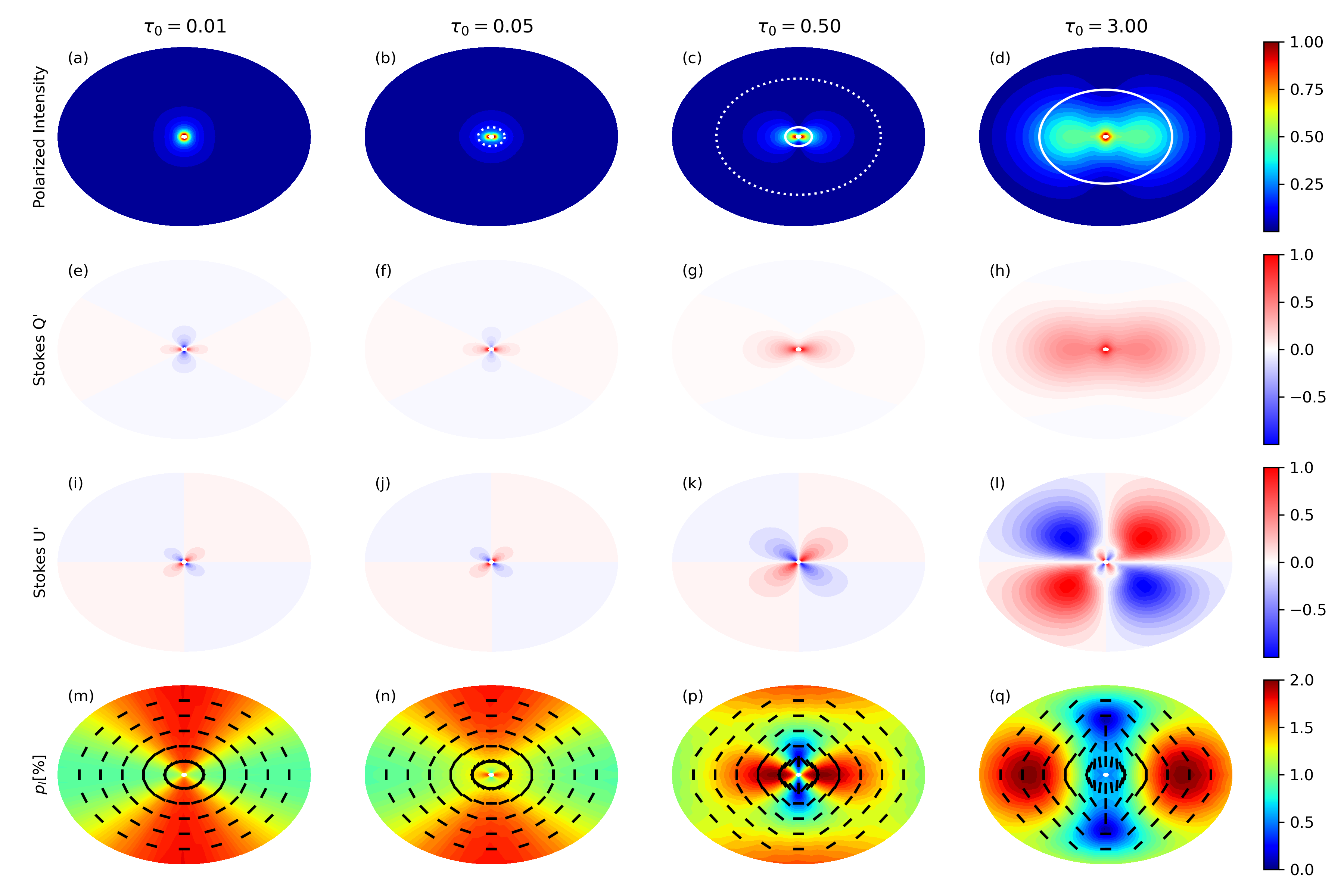}
    \caption{
        The reference model images pieced together by plane-parallel slab calculations. The columns correspond to different $\tau_{0}$, and from the left to right, the optical depth increases. The rows from top to bottom are the linear polarized intensity, Stokes $Q'$, Stokes $U'$, and the linear polarization fraction. The color maps for the first three rows are shown relative to its own peak, while $p_{l}$ is the actual level of polarization in percent. The vectors in the last row denote the polarization angle, and the vector lengths are uniform. The white dotted contour in the first row shows where the total optical depth along the line of sight is 0.1, and the white solid contour shows where the total optical depth is 1. 
    }
    \label{fig:model_tau0}
\end{figure*}

Fig. \ref{fig:model_tau0} shows the linear polarized intensity, Stokes $Q'$, Stokes $U'$, and $p_{l}$ along with the polarization vectors at different $\tau_{0}$. The different $\tau_{0}$ values are chosen to illustrate particular changes of features in the polarization images. 

In the most optically thin case, $\tau_{0}=0.01$, the polarization fraction is largest along the minor axis and smallest along the major axis of the disk (Fig. \ref{fig:model_tau0}m). Also, the polarization vectors are elliptical. At this limit, scattering is negligible, and thus the polarization pattern is a simple result of the viewing geometry of the azimuthally aligned prolate grains. Along the minor axis of the disk, the symmetry axis of the grain is directed horizontally and seen from the longest side which produces the largest thermal polarization. The Stokes $Q'$ is negative and the Stokes $U'$ is zero because the polarization is completely horizontal. This corresponds to the $\phi=90^{\circ}$ case in Section \ref{sec:plane_parallel}. The grains along the major axis are seen more pole-on, and thus the projection decreases the thermal polarization. Since the polarization is completely vertical, the Stokes $Q'$ is positive, and Stokes $U'$ is zero. In Fig. \ref{fig:model_tau0}, the polarized intensity is the greatest near the center simply because the temperature is highest. Additionally, it is slightly larger along the minor axis than along the major axis for a given radius because the polarization fraction is higher (this is more obvious when beam convolution is considered in Section~\ref{ssec:model_tau0_conv}). 

With a slight increase in optical depth, $\tau_{0}=0.05$, the outer regions of the disk remains largely similar to the most optically thin case across all four polarization properties. However, the inner regions in the polarization fraction (Fig.~\ref{fig:model_tau0}n) begin to see a horizontal bar of higher polarization fraction. This is due to the emergence of a positive Stokes $Q'$ as we can see in Fig.~\ref{fig:model_tau0}f which is a result of scattering. 

The trend continues to the $\tau_{0}=0.5$ case, where the positive Stokes $Q'$ becomes even more obvious. The polarized intensity in Fig.~\ref{fig:model_tau0}c, the Stokes $Q'$ in Fig.~\ref{fig:model_tau0}g, and $p_{l}$ in Fig.~\ref{fig:model_tau0}p all have two prominent lobes across the major axis. The lobes are formed because the thermal polarization and inclination-induced polarization add with each other. In addition, $p_{l}$ begins to develop two polarization ``holes'' along the minor axis above and below center. This corresponds to the location where Stokes $Q'$ changes from a positive value to a negative value going away from the center, in which case the polarization vector shifts by $90^{\circ}$ in Fig.~\ref{fig:model_tau0}p (see also $\phi=90^{\circ}$ of Fig.~\ref{fig:q_opt}). The pattern of the polarization fraction in the outer regions of the $\tau_{0}=0.01, 0.05$ cases (Fig. \ref{fig:model_tau0}m, n) disappears as thermal polarization gives way to scattering polarization due to higher optical depth for $\tau_{0}=0.5$ and $3$.

\subsubsection{Effects of Finite Beam Size} \label{ssec:model_tau0_conv}

\begin{figure*}
    \centering
    \includegraphics[width=\textwidth]{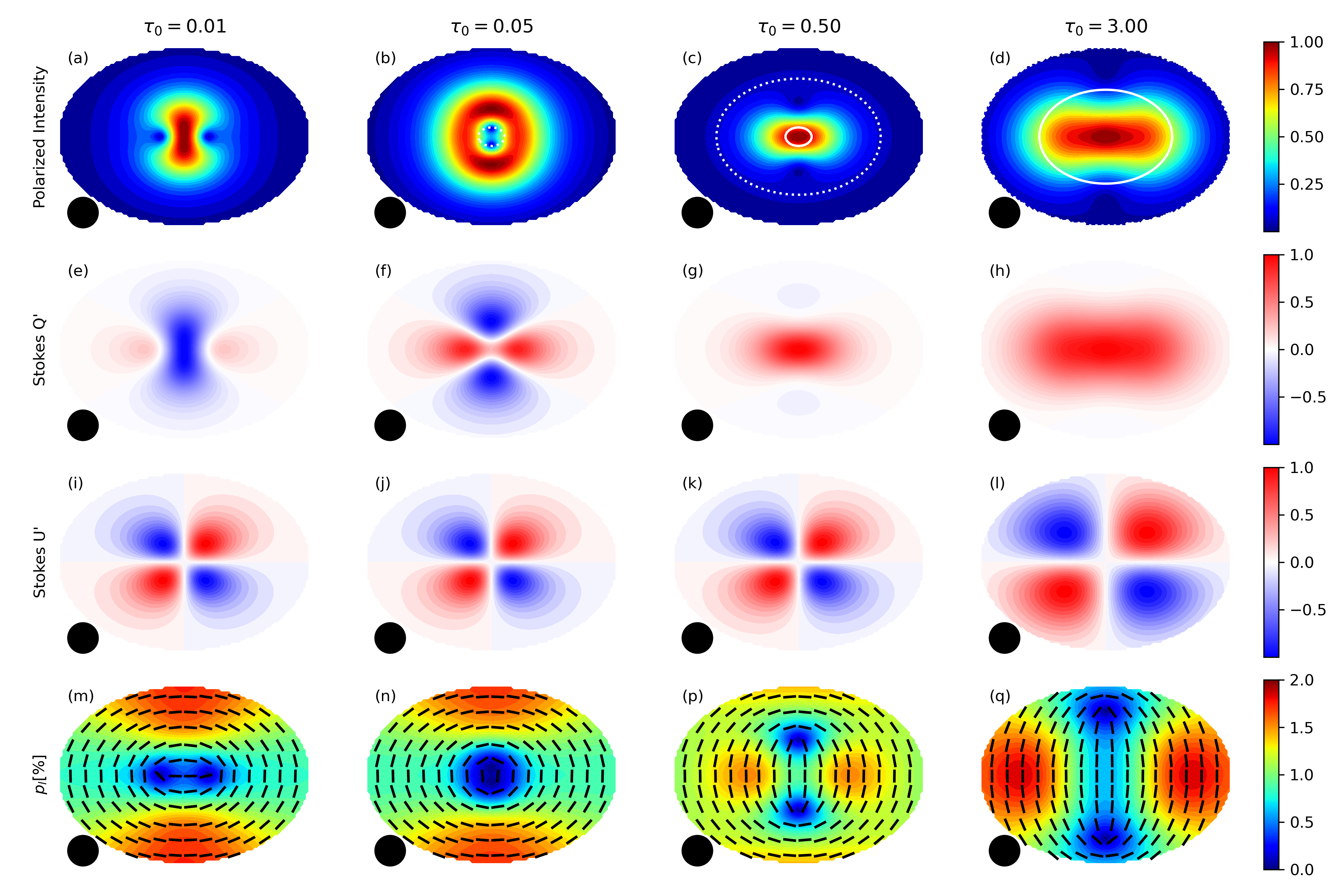}
    \caption{
        The same reference model images but including beam convolution. The figure is plotted in the same way as Fig. \ref{fig:model_tau0}, except the black circles show the beam size. 
    }
    \label{fig:model_tau0_conv}
\end{figure*}

One qualitative feature that is not captured in the model is the low polarization region at the center of Band 3 seen in Fig.~\ref{fig:obs_rot}a and j. We demonstrate that this is due to beam convolution. In Fig.~\ref{fig:model_tau0_conv}, we show the same model disk but convolved to a circular Gaussian beam with FWHM of 50 au. 

When the optical depth is low, $\tau_{0}=0.01$, the polarization at the center of the disk can be horizontal. The polarization along the minor axis is larger since the prolate grains are viewed from the edge (corresponding to a large negative Stokes $Q'$), while along the major axis, the prolate grain is viewed closer to its pole resulting in a smaller polarization (corresponding to a small positive Stokes $Q'$). Beam averaging leaves more negative Stokes $Q'$ as a result. Also, the Stokes $U'$ is symmetric around the center, which averages out to zero. Therefore, the resulting polarization direction is completely horizontal. 

We see that in the case of $\tau_{0}=0.05$, the polarized intensity and $p_{l}$ are low in the center in Fig.~\ref{fig:model_tau0_conv}b, n. This is because the increase in $\tau_{0}$ allows scattering to produce more positive Stokes $Q'$ particularly in the central region of the disk where the optical depth is higher. The extra positive Stokes $Q'$ can compete with the largely negative Stokes $Q'$ just from thermal polarization. Beam averaging mixes the two components which cancel with each other. For the $\tau_{0}=0.5$ and $3$ cases, the polarization vectors around the disk center are parallel to the disk minor axis as positive Stokes $Q'$ from scattering dominates.

\subsection{Folding Images} \label{ssec:fold_image}
There are two symmetries that we can exploit when assuming azimuthally aligned prolate grains and a geometrically thin disk. For any inclined axisymmetric disk, the image is mirror symmetric across the disk minor axis (i.e., the left half is mirror symmetric to the right half in the principle frame).\footnote{Geometrically thick disks or highly inclined disks can feature near/far-side asymmetry in the Stokes $I'$ and polarization properties \citep{Yang2017}, but the image will remain mirror symmetric across the minor axis.} For Stokes $U'$, the symmetry involves a change in sign. Since we further assumed that the disk is geometrically thin, the image is also mirror symmetric across the disk major axis (the top half to the bottom half) with a change in sign for Stokes $U'$ again. The models in Fig. \ref{fig:model_tau0} clearly demonstrate the two symmetries.

The expected symmetries for an axisymmetric disk motivates us to ``fold'' the observed images twice: we first average $\vect{I}'$ across the disk minor axis and then average across the disk major axis. The benefits of folding is to average out the asymmetries and to improve the signal-to-noise ratio of the data. The asymmetries in the observed data (Fig.~\ref{fig:obs_rot}) cannot be explained by the axisymmetric disk model adopted in this paper and in most of the disk polarization models in the literature to date. Thus, we opt to remove the asymmetries for comparison with axisymmetric disk models as a first step towards a more comprehensive model. Averaging across the two symmetries increases the signal-to-noise by a factor of 2. Table \ref{tab:snr_improvement} lists the improvement of the signal-to-noise for each Stokes parameter at each wavelength (before the image was corrected by the primary beam). We adopt a factor of 2 improvement for simplicity in the following analysis. 

We should caution that an elongated observing beam can break the two symmetries described above. If the beam is elongated and not along the disk major or minor axis, then the image does not have any symmetry. The beams of the Bands 3 and 7 images are not along the disk major or minor axis, but they are at least roughly circular. The Band 6 beam happens to be elongated roughly along the disk major axis, which should alleviate the impact of the beam orientation. Utilizing the symmetries in the visibility domain may produce more robust results, but it is beyond the scope of this paper. 

\begin{table}
    \centering
    \begin{tabular}{c|c|c|c}
        Band & $I$ & $Q$ & $U$ \\
        3 & 1.5 & 1.9 & 2.0 \\
        6 & 1.9 & 3.2 & 1.2 \\
        7 & 2.0 & 2.1 & 2.8 
    \end{tabular}
    \caption{  
        Empirically measured improvement in the signal-to-noise for the Stokes parameters at each wavelength after image folding. 
    }
    \label{tab:snr_improvement}
\end{table}
%% V: 2.1, 1.9, 1.9

\begin{figure*}
    \centering
    \includegraphics[width=\textwidth]{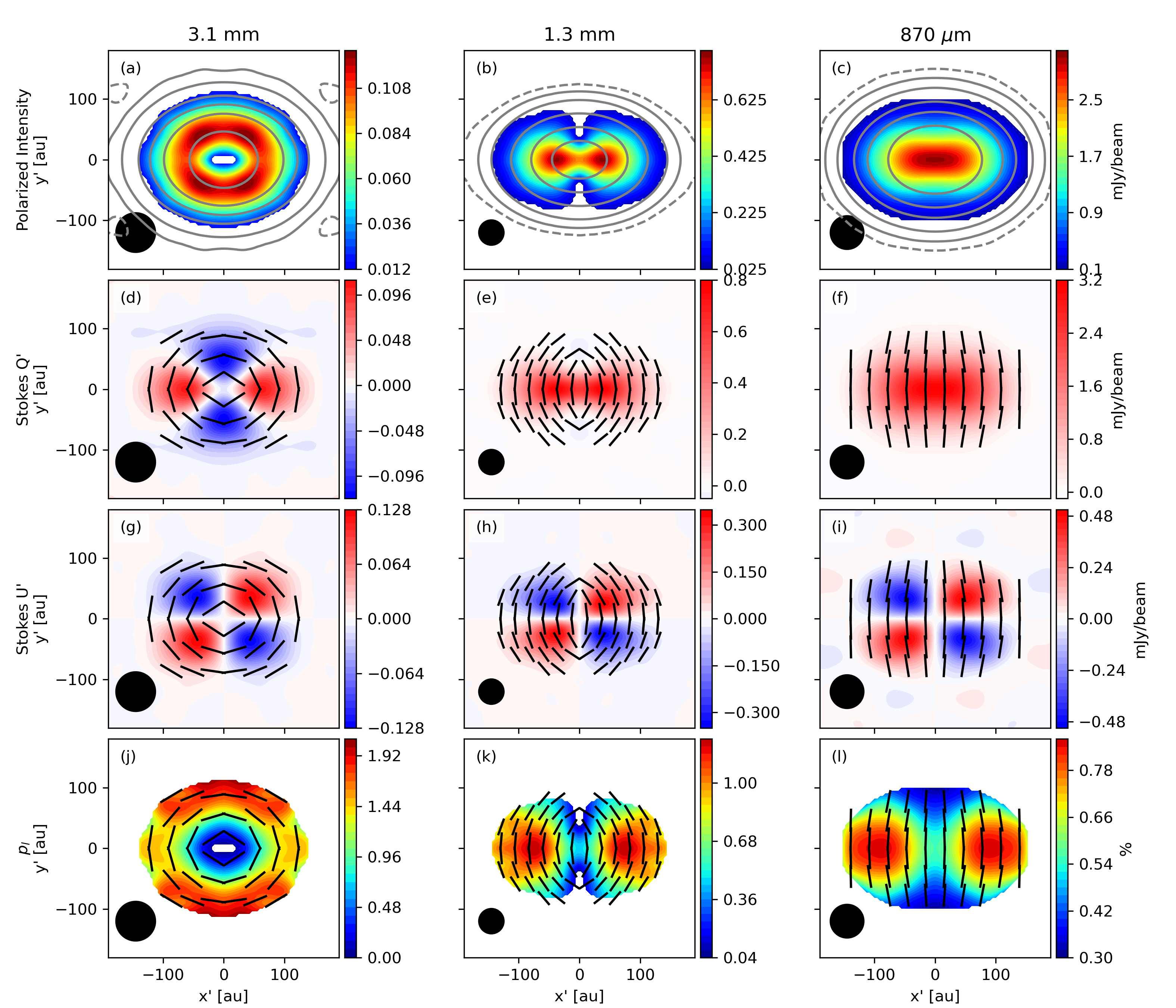}
    \caption{
        The results after folding the images along the major and minor axes. The figure is plotted in the same way as Fig. \ref{fig:obs_rot}. 
    }
    \label{fig:obs_rot_fold}
\end{figure*}

Fig. \ref{fig:obs_rot_fold} shows the folded images for all three wavelengths. After folding the image twice, we copy the quadrant back into the four quarters for visualization. We take the geometric mean of the beam major and minor axis as a rough estimate of the resolution. One benefit from folding is that the azimuthal variation of the $p_{l}$ at Band 3 becomes clearer. In particular, the level of $p_{l}$ is greater along the minor axis than along the major axis which is what we expect from azimuthally aligned prolate grains (see Fig.~\ref{fig:model_tau0}m, n and Fig.~\ref{fig:model_tau0_conv}m, n) if the image is optically thin\footnote{In fact, \cite{Carrasco2019} inferred the Band 4 optical depth to be $\sim 1$ from $\sim 60$ to $100$ au which suggests that Band 3 is optically thin over most of the disk where polarization is detected.}. 

\subsection{Decomposition of Observed Polarization into Scattering and Thermal Components} \label{ssec:empirical_pol_azi}
In Section \ref{sec:plane_parallel}, we treated self-scattering of aligned grains consistently. From the assumed aspect ratio of the grain, we can observe that, to a good approximation, the thermal polarization of non-spherical grains adds linearly with scattering polarization of spherical grains to produce the polarization of scattering aligned grains (see Fig.~\ref{fig:azi_prof_taum}). Furthermore, assuming azimuthally aligned grains, the thermal polarization varies with azimuth, while scattering polarization of spherical grains is constant across azimuth. Thermal polarization adds with scattering polarization along the major axis of the disk. On the other hand, along the minor axis of the disk, thermal polarization cancels with scattering polarization. The superposition breaks down for optically thick cases (which is more likely for Band 7), but the error is regulated since we have the least contribution from thermal polarization. Building on these two approximations, we can roughly differentiate the level of polarization from thermal emission and that from scattering from a single image.

Assuming the grain is in the dipole limit, the thermal polarization depends on the projected shape of the grain. For a prolate grain, we have \citep{Lee1985, Yang2016_oblate}:
\begin{align} \label{eq:p_theta_g}
    p(\theta_{g} ) = \dfrac{ p_{0} \sin^{2} \theta_{g} }{1 - p_{0} \cos^{2} \theta_{g} } \sim p_{0} \sin^{2} \theta_{g}
\end{align}
where $\theta_{g}$ is the viewing angle from the axis of symmetry of the prolate grain. If $\theta_{g}=0^{\circ}$, the grain is seen pole-on, while $\theta_{g}=90^{\circ}$ means the grain is seen from the side (perpendicular to the axis of symmetry). We call $p_{0}$ the intrinsic polarization, since it is the maximum polarization of the prolate grain determined just from its shape. The approximation in the right-hand-side applies because $p_{0}$ is much less than unity (recall Band 3 polarization is only $\sim2\%$). 

The polarization is attenuated by dichroic extinction as optical depth increases \citep{Hildebrand2000}. However, since $p_{0}$ is small, the optical depth attenuates the polarization by the same factor regardless of $\theta_{g}$. Thus, we define the $p_{0}$ attenuated by optical depth as $T_{0}$; the ``$T$'' stands for the ``thermal'' polarization component (not to be confused with the temperature $T$). The observed thermal polarization after attenuation as a function of $\theta_{g}$ is 
\begin{align} \label{eq:T_theta_g}
    T_{p}(\theta_{g}) = T_{0} \sin^{2} \theta_{g} \text{ .}
\end{align}

Along the minor axis of the disk, the azimuthally aligned prolate grain produces a negative Stokes $Q'$ since the projected elongation is parallel to the disk major axis. The magnitude of the polarization is just $T_{0}$, since the grain is seen from the edge (or $\theta_{g}=90^{\circ}$). Thus, thermal polarization cancels with the polarization due to scattering which makes: 
\begin{align} \label{eq:q_minor}
    q'_{\text{minor}} = S - T_{0}
\end{align} 
where $S$ is the polarization fraction due to scattering approximated by spherical grains.
Along the major axis of the disk, $\theta_{g}=90^{\circ} - i$ and the thermal polarization has the same sign as scattering polarization. Thus, the polarization due to scattering and thermal polarization add together:
\begin{align} \label{eq:q_major}
    q'_{\text{major}} = S + T_{0} \cos^{2} i \text{ .}
\end{align}
From the two algebraic equations, we can easily solve for the two unknowns $S$ and $T_{0}$.\footnote{We should note that scattering aligned grains can produce Stokes V which is not captured in the simple linear combination. Given that HL Tau currently does not have a firm detection of Stokes V, the available data does not contradict the linear combination approximation. }

\begin{figure}
    \centering
    \includegraphics[width=\columnwidth]{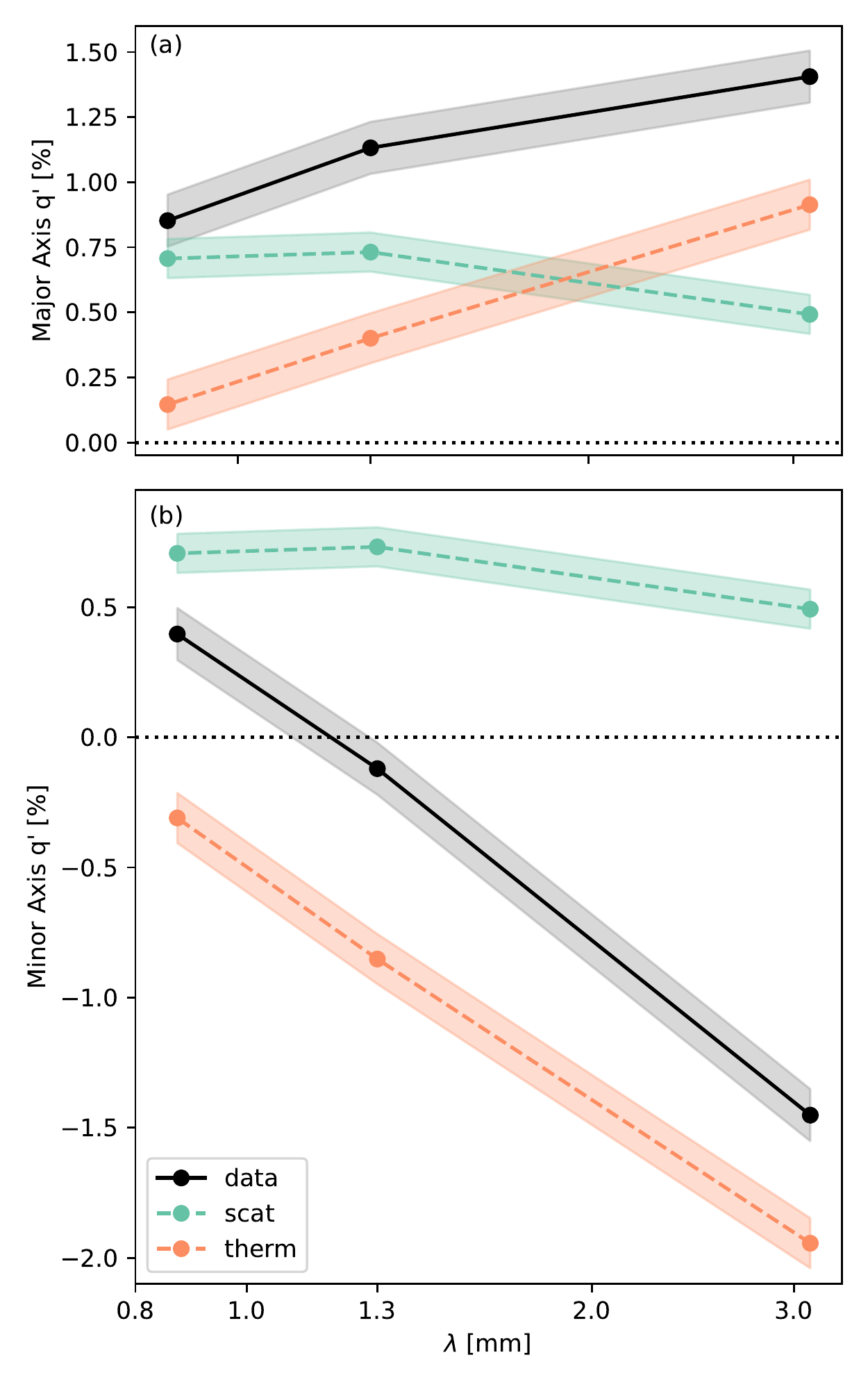}
    \caption{
        The spectrum of $q'$ together with the inferred contributions from thermal emission and scattering measured at a representative radius of $100$ au. The top panel is from the major axis, while the bottom panel is from the minor axis. The black lines come from the HL Tau data, and its shaded regions is the uncertainty based on the statistical noise level. The green lines are the contribution to the observed polarization from scattering $S$, and the orange lines are that from thermal polarization, which is $T_0 \cos  i^2$ on the major axis (upper panel) and $-T_0$ on the minor axis (lower panel). Their shaded regions are their uncertainties. The horizontal dotted line is $q'=0$ for guidance. 
    }
    \label{fig:q_spectrum}
\end{figure}

In Fig. \ref{fig:q_spectrum}, we show the observed $q'_{\text{major}}$ and $q'_{\text{minor}}$ across wavelength at a radius of 100 au. The radius is chosen to maximize the number of independent beams across azimuth, but also to avoid low signal-to-noise. The uncertainties are based on statistical error from the noise levels of the folded Stokes $Q'$ and Stokes $I'$. For the rest of the paper, we only consider statistical noise, but for noise levels that are less than $0.1\%$, we use $0.1\%$ since the ALMA instrumental error of polarization fraction is expected to be about $0.1\%$. From just the observational data, it is easy to see why scattering polarization alone does not work and thermal polarization is necessary. Based on $q'_{\text{major}}$ (Fig. \ref{fig:q_spectrum}a), the measured polarization fraction increases with increasing wavelength. This is opposite to what is expected for scattering alone, which should be a near monotonic decrease \citep{Lin2020_specpol}.

Furthermore, we also plot the disentangled polarization due to scattering $S$ and that due to thermal polarization, which is $T_{0} \cos^{2} i$ along the major axis and $-T_{0}$ along the minor axis. The uncertainties are based on error propagation. We can see that $S$ is smaller at Band 3 than at Band 6 and 7, while the $S$ between the Bands 6 and 7 are comparable. The behavior is expected from scattering. As seen in Fig.~\ref{fig:q_opt}, the scattering polarization of spherical grains increases with increasing optical depth until the optical depth is $\sim 1$. Just from the observed spectrum, we can infer that Band 3 is optically thin, while Band 6 and 7 have optical depths of unity or larger.

The contribution from thermal polarization is the least at Band 7 and gradually increases to Band 3. The monotonic increase in thermal polarization to longer wavelength is also expected because the optical depth decreases which leads to less dichroic extinction as depicted by the non-scattering aligned grains case in Fig. \ref{fig:q_opt}.

Once we solved for $T_{0}$, we can predict the thermal polarization at any azimuth based on geometrical arguments. Recall that $\theta_{g}$ is the viewing angle of the grain or the angle from the axis of symmetry ($x$-axis) to the observer. Since we assume the grain is azimuthally aligned in the midplane of the disk, $\theta_{g}$ simply depends on the inclination and azimuth of the disk. We can obtain the relation from geometrical arguments: 
\begin{align} \label{eq:theta_g}
    \cos \theta_{g} = \sin i \cos \phi_{d} 
\end{align}
which gives the polarization in the grain frame through Eq.~(\ref{eq:T_theta_g}). Rotating into the lab frame will give us $q'$ and $u'$ from both scattering and thermal emission:
\begin{align} 
    q' &= S + T_{0} ( \cos^{2} i \cos^{2} \phi_{d} - \sin^{2} \phi_{d} ) \label{eq:rotated_analytical_q} \\
    u' &= - T_{0} \cos i \sin 2 \phi_{d} \label{eq:rotated_analytical_u}
\end{align}
(see Appendix \ref{sec:lab_to_grain_frame} for details).

\begin{figure*}
    \centering
    \includegraphics[width=\textwidth]{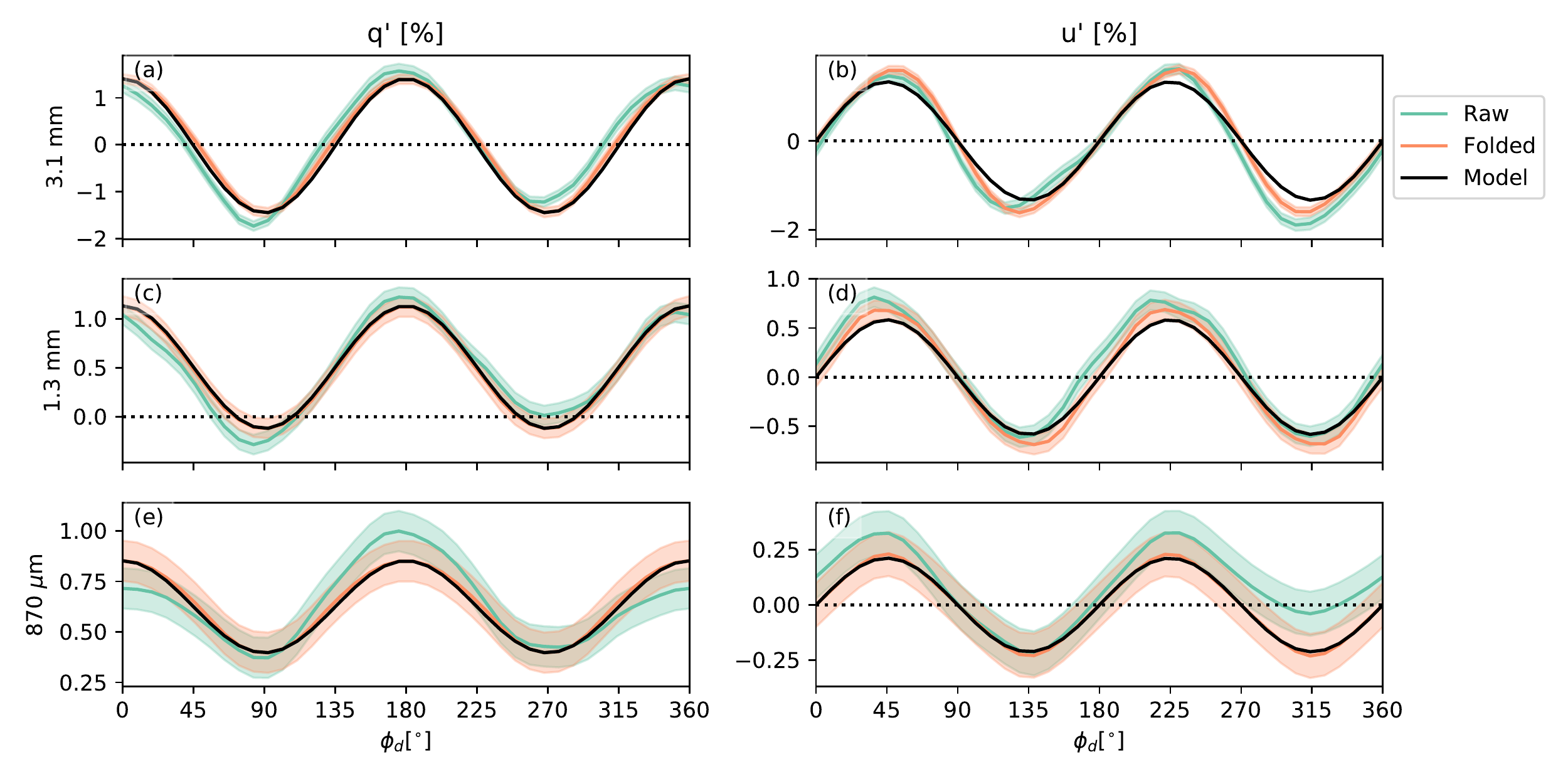}
    \caption{
        The azimuthal profile of $q'$ and $u'$ in percent from the raw data, the folded data, and the empirical model in green, orange, and black, respectively. The shaded regions are their noise uncertainties. $\phi_{d}=0^{\circ}$ and $180^{\circ}$ correspond to the disk major axis and $\phi_{d}=90^{\circ}$ and $270^{\circ}$ correspond to the disk minor axis. 
    }
    \label{fig:qu_azi}
\end{figure*}

In Fig.~\ref{fig:qu_azi}, we compare the predicted azimuthal profiles for $q'$ and $u'$ based on the values of $S$ and $T_0$ determined from the polarization data on the major and minor axes (and shown in Fig.~\ref{fig:q_spectrum}) with the observed azimuthal variations. Although there are some discrepancies (e.g., around $\phi_{d}\sim 80^{\circ}$ for $q'$ in Band 3 and $\sim 45^{\circ}$ for $u'$ in Bands 3 and 6), the predicted profiles match the observed ones, especially the folded data, remarkable well overall, which adds support to our empirical method of decomposing the observed polarization into scattering and thermal components.

\section{Discussion} \label{sec:discussion}

\subsection{Reconstruction of the Polarization Images}

In Section \ref{ssec:empirical_pol_azi}, we obtained an empirical measurement of scattering polarization $S$ and thermal polarization $T_{0}$ for one radius. In principle, one can measure $S$ and $T_{0}$ at each point across radius and infer the property of grains as a function of radius as the resolution may allow. However, for the HL Tau data, there is roughly only one beam across the disk minor axis, which limits this empirical technique to just one independent data point at $\sim 100$au. 

\begin{figure}
    \centering
    \includegraphics[width=\columnwidth]{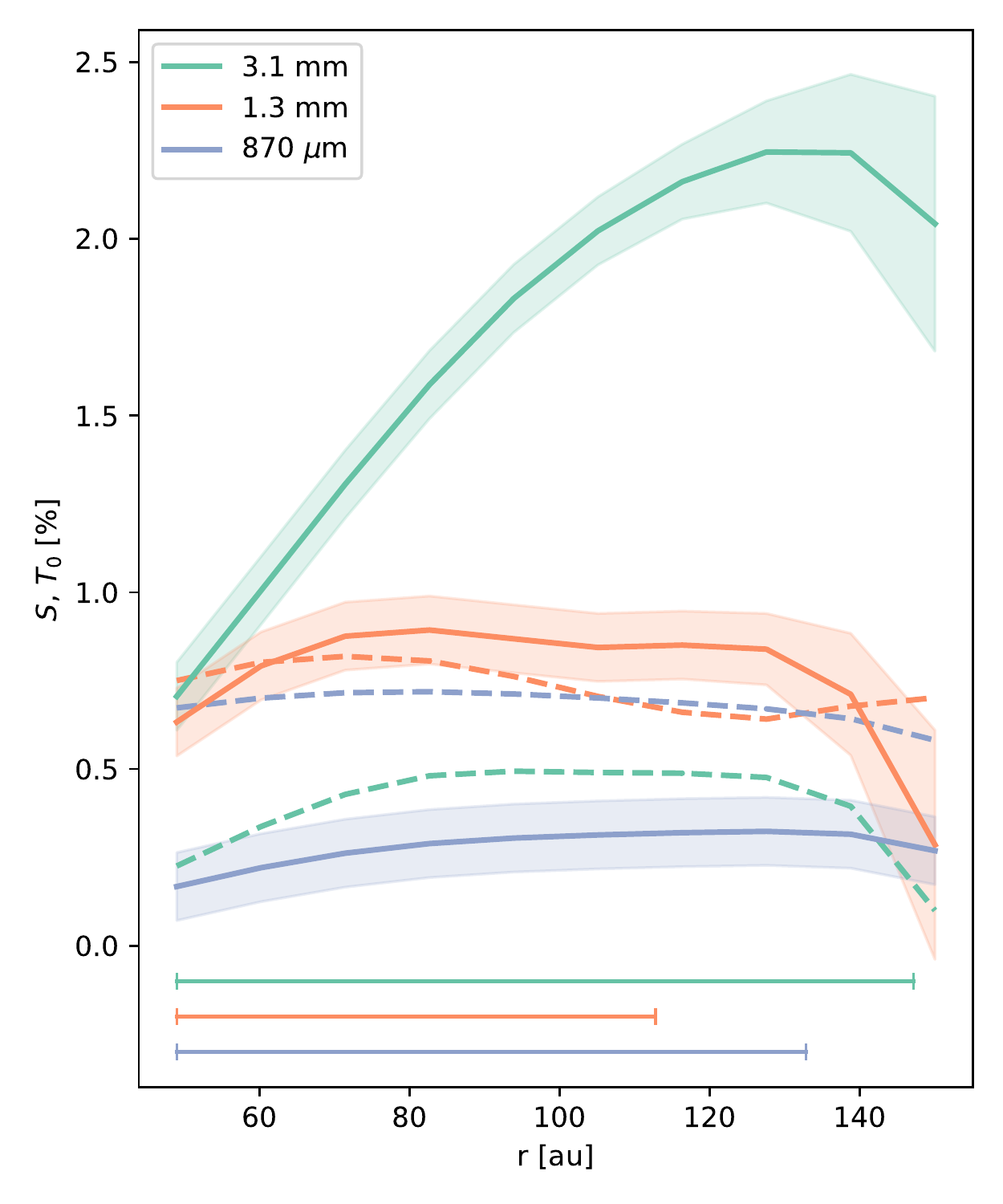}
    \caption{
        The radial profile of $S$ (dashed lines) and $T_{0}$ (solid lines) at each wavelength. The colors, green, orange, and indigo, correspond to Band 3, 6, and 7 respectively. Only the $T_{0}$ profiles are shaded to show their noise uncertainties for readability, but the uncertainty for $S$ are comparable to that of $T_{0}$. The horizontal line segments at the bottom represent the length of the beam projected onto the minor axis of the disk in colors of the respective wavelength.
    }
    \label{fig:ST0_radial}
\end{figure}

Nevertheless, as a consistency check, we solve for $S$ and $T_{0}$ at different radii in a single image under the influence of the finite beam size. We choose an inner radius of $50$au which gives $\sim4$ beams around the azimuth of the Band 3 image and the outer radius is limited by sensitivity. In Fig.~\ref{fig:ST0_radial} we show the radial profile at each wavelength. As a comparison, we show the beam sizes projected along the disk minor axis, FWHM$/ \cos i$, since we are mainly limited by the resolution along the minor axis for independent radial points.

We find that $S$ is roughly constant of radius for each wavelength. $T_{0}$ decreases towards the inner radius likely due to beam convolution which can average out the azimuthal variation (see Fig.~\ref{fig:model_tau0_conv}). Similar to Fig.~\ref{fig:q_spectrum}, one can see immediately that the thermal component $T_{0}$ dominates the scattering component $S$ everywhere for Band 3 (the green solid and dashed lines in Fig.~\ref{fig:ST0_radial}), especially at large radii where effects of beam averaging is less. The opposite is true for Band 7 with larger $S$ than $T_{0}$ and at Band 6, the two components are comparable. 

\begin{figure*}
    \centering
    \includegraphics[width=\textwidth]{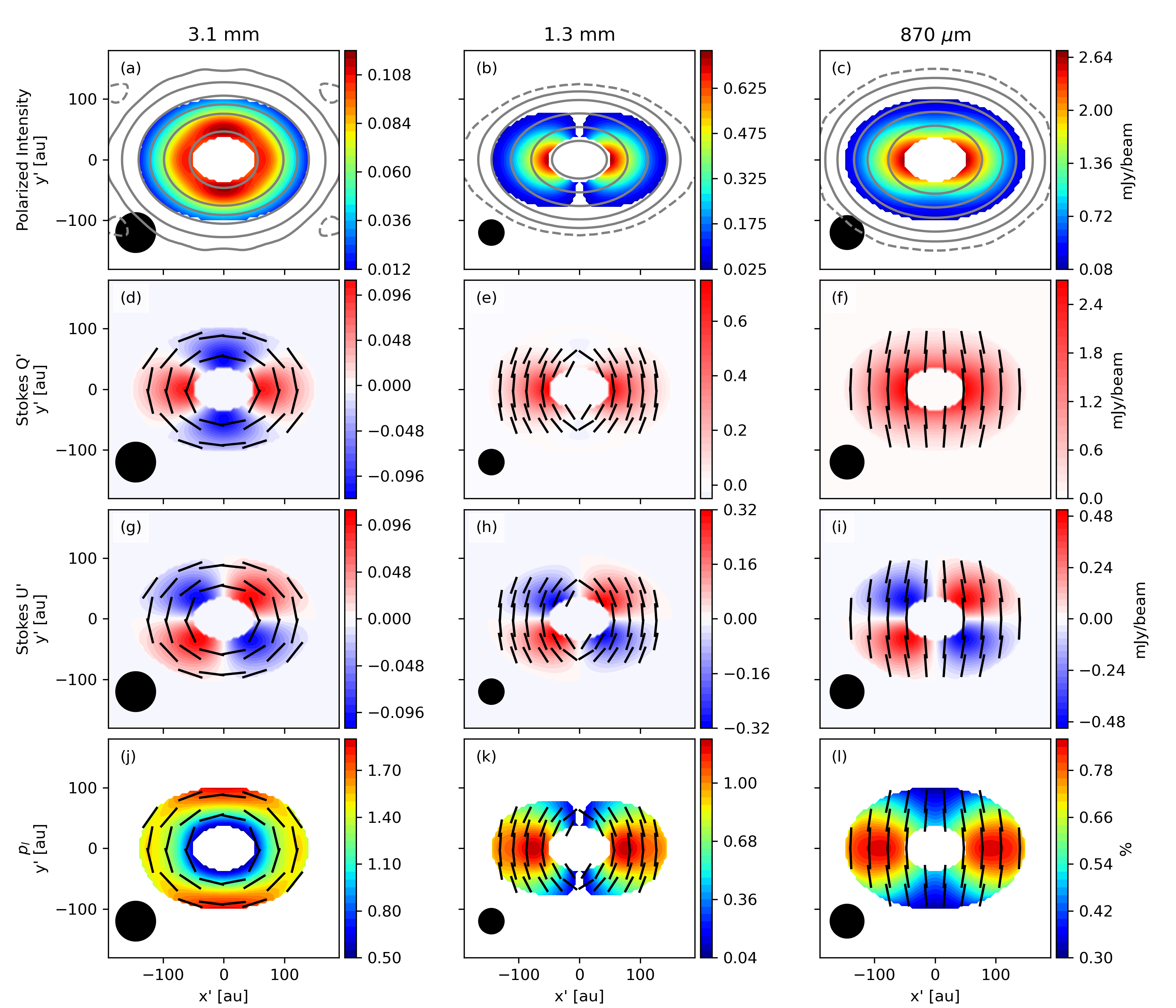}
    \caption{
        The reconstructed polarization images using the empirical model. The figure is plotted in the same way as Fig.~\ref{fig:obs_rot} and is to be compared with Fig.~\ref{fig:obs_rot_fold} that displays the observed multiwavelength polarization images after folding. 
    }
    \label{fig:reconstructed_emp}
\end{figure*}

From Section \ref{ssec:empirical_pol_azi}, we predicted the azimuthal profiles of $q'$ and $u'$ based on $S$ and $T_{0}$. Since we can solve the azimuthal profile at each radius, we can produce an image of $q'$ and $u'$. To retrieve Stokes $Q'$ and $U'$, we simply multiply the solved $q'$ and $u'$ by the observed Stokes $I'$ (Eq.~(\ref{eq:q_u})). In Fig. \ref{fig:reconstructed_emp}, we show the reconstructed polarization images plotted in a similar manner as the folded images in Fig. \ref{fig:obs_rot_fold}. The hole in the reconstructed images are simply because of our inner cut of radius. Indeed, the polarization images are quantitatively similar to the folded data. 

One notable difference is the polarized intensity at Band 3. In Fig. \ref{fig:obs_rot_fold}a, the polarized intensity has four peaks instead of expected two peaks along the minor axis in Fig. \ref{fig:reconstructed_emp}a. This is likely because of the intrinsic asymmetry at Band 3. Shown in Fig. \ref{fig:obs_rot}a, there is a very strong asymmetric peak of the polarized intensity that is not along the minor axis. Even folding the data could not cancel out the asymmetry which leaves a footprint in the folded image as a bright spot slightly offset from the minor axis. Since the folded image is symmetric across the disk major and minor axes, the bright spot is imprinted across the four quadrants. The cause of the intrinsic asymmetry is unclear. 

Despite the above difference, the simple analytic decomposition of the observed polarization into the scattering and thermal components (Fig.~\ref{fig:ST0_radial}) and the predicted multiwavelength polarization distributions based on the decomposition and simple geometric effects match the folded observational data remarkable well (compare Fig.~\ref{fig:reconstructed_emp} and Fig.~\ref{fig:obs_rot_fold}). In particular, as the observing frequency increases, there is a transition from a more azimuthal pattern to a more  pattern along the minor axis in the polarization orientation and from an azimuthal distribution of polarization fraction that is higher along the minor axis than along the major axis to the opposite (i.e., the dumbbell-shaped  distribution along the major axis) in both the observation data and the model prediction. Fig.~\ref{fig:ST0_radial} quantified the degree to which the Band 3 polarization is dominated by thermal emission and the Band 7 polarization is dominated by scattering. It shows that the crossover occurs near Band 6, where the thermal and scattering components are comparable, which is consistent with the conclusion of \cite{Stephens2017} based on a simple mix of a uniform and an azimuthal polarization patterns. 

The analytic decomposition based purely on the observational data and geometry without prior knowledge of the disk or dust properties (except for the assumptions of axisymmetry and the geometrically thin limit) demonstrated in Section \ref{ssec:empirical_pol_azi} can be a unique tool for analyzing resolved polarization images. Plane-parallel slab calculations have been used to fit the radial properties of disks across wavelength without assuming any power-law if the multiwavelength images are resolved \citep[e.g.][]{Carrasco2019, Yen2020, Sierra2021}. If polarization images can be resolved across wavelength, the empirical decomposition method or plane-parallel slab calculations can be used to determine even more properties of grains, such as the degree of alignment or scattering properties of the grain.

\subsection{Properties of the Grain from the Polarization Spectrum}

\subsubsection{Aligned Grains}
In Section \ref{ssec:empirical_pol_azi}, we derived empirical constraints on the scattering and thermal components of the observed polarization, $S$ and $T_{0}$, at $100$ au. From $T_{0}$, we can estimate a lower limit to the aspect ratio of the grains if they are perfectly aligned optically thin and in the dipole limit. In the coordinate system with the axes along the three principle axes of the grain, the polarizability matrix is diagonal and the polarizability with respect to axis $i$ are \citep{Bohren1983}:
\begin{align}
    \alpha_{i} = V \dfrac{m^{2} - 1}{1 + L_{i}(m^{2} - 1) } \text{ , } i=1,2,3
\end{align}
where $V$ is the volume of the ellipsoid and $m$ is the complex refractive index. $L_{i}$ are the geometrical parameters along each axis $i$ that satisfy $\sum L_{i}=1$. The expression for $L_{1}$ is \citep{Bohren1983}:
\begin{align}
    L_{1} = \dfrac{1 - e^{2} }{ e^{2} } \bigg( -1 + \dfrac{1}{2e} \ln \dfrac{1 + e}{1 - e} \bigg) 
\end{align}
where $e^{2} \equiv 1 - s^{2}$ and $s$ is the aspect ratio (Section \ref{sec:plane_parallel}). From the optical theorem, we can get the absorption opacity along the three principle axes of the prolate grain
\begin{align}
    C_{\text{abs},i} = \dfrac{2 \pi }{\lambda } \operatorname{Im}( \alpha_{i} ) \text{.}
\end{align}
The intrinsic polarization of the grain in Eq. (\ref{eq:p_theta_g}) is 
\begin{align} \label{eq:p0_dipole}
    p_{0} = \dfrac{ C_{\text{abs},1} - C_{\text{abs},2} }{ C_{\text{abs},1} +  C_{\text{abs},2}}
\end{align}
when the grain axis of symmetry is perpendicular to the line of sight. 

Since $p_0\sim T_0\sim 2\%$ for Band 3 at 100 au (Fig. \ref{fig:q_spectrum}) and we can expect that Band 3 is optically thin \citep{Carrasco2019}, we can obtain an upper limit to the aspect ratio of the grain of $\sim 0.97$ if the grains are also perfectly aligned. Grains that are more spherical than that cannot produce the required amount of thermal polarization. In the case of poor alignment, grains more elongated than $\sim 0.97$ could produce similar levels of thermal polarization. Also, porosity could allow elongated grains to maintain low $p_{0}$ \citep{Kirchschlager2019}. The implications will require more exploration. 

% perhaps mention Kirchschlager+2019 on elongated grains with porosity

\subsubsection{Scattering Polarization}

In addition, we can estimate the grain size based on the level of polarization due to scattering $S$. Using the plane-parallel slab, we calculate the polarization, $q$, varying in grain size and density, but with only spherical grains since $S$ is an approximation of the level of scattering for spherical grains (as motivated by Section \ref{sec:plane_parallel}). Since $S$ includes the effects of beam size, we convolve the three images to a common beam of 0.51$\arcsec$ to take out any systematic differences due to resolution for fitting. 

\begin{figure}
    \centering
    \includegraphics[width=\columnwidth]{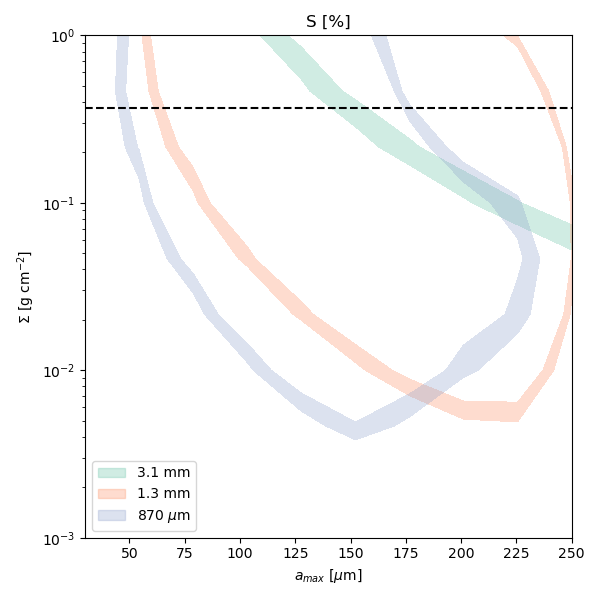}
    \caption{
        The polarization fraction in the parameter space of surface density and grain size. The shaded regions correspond to the allowed parameter space given the observed $S$ and its $1\sigma$ uncertainties. The black horizontal dashed line is the expected dust surface density that would make the disk graviationally unstable with $Q=1$. 
    }
    \label{fig:sph_S}
\end{figure}

In Fig.~\ref{fig:sph_S}, we use shaded areas to denote regions in the parameter space that is consistent with the the empirically derived values of $S$ for the three ALMA Bands to within $1\sigma$ uncertainty. We calculate the slab at the three wavelengths and any intersecting regions should reveal a solution. We only consider grain sizes up to $250 \mu$m because larger grain sizes easily produce negative polarization at Band 7, i.e., the polarization direction becomes perpendicular to the disk minor axis, which is inconsistent with the observations data \citep[e.g.][]{Yang2016_inc, Kataoka2016_hltau, Yang2020}. The maximum dust surface density $\Sigma$ considered is 1 g cm$^{-2}$ motivated by the modeling results from \cite{Pinte2016} and from computational constraints. Nevertheless, the maximum dust surface density is already gravitationally unstable. Consider the Toomre $Q$ parameter
\begin{align}
    Q = \dfrac{c_{s} \Omega }{ \pi G \Sigma_{g} }
\end{align}
where $c_{s}$ is the sound speed, $\Omega$ is the Keplerian angular velocity, and $\Sigma_{g}$ is the gas surface density. At a radius of $100$ au, the temperature is $\sim 20$ K \citep{Okuzumi2016, Pinte2016}. The stellar mass of HL Tau is $2.1 \pm 0.2 M_{\odot}$ \citep{Yen2019}. Gravitationally stable disks should have $Q > 1$, and thus we use $Q=1$ to obtain an upper limit to the gas surface density. We assume the typical dust-to-gas mass ratio of 0.01 to obtain the dust surface density of $\sim 0.4$ g cm$^{-2}$. Thus, we do not expect a reasonable solution beyond 1 g cm$^{-1}$ under typical assumptions. 

For Band 6 and 7 ($\lambda=1.3$mm and $870\mu$m), the shaded region forms a loop in the parameter space which is understandable. Iterating along $\amax$, the polarization peaks when the size parameter $x$ is near unity, while along $\Sigma$, the polarization peaks when the optical depth is of order unity as demonstrated in Fig.~\ref{fig:q_opt}. Outside the loop, the polarization is too low compared to the observations, while the polarization is too high within the loop. Thus, the loop essentially traces around the polarization ``hill top'' in the parameter space. For Bands 3, we only see a part of the loop; its ``hill top'' is towards the upper right.

We find that there is no parameter space in $\Sigma$ and $\amax$ that can simultaneously explain the polarization fraction $S$ at all three wavelengths. Interestingly, the empirically derived $S$ of $\sim 0.5\%$ at Band 3 is similar to the level of polarization from self-scattering from 130 $\mu$m grains as determined in \cite{Mori2021}. The solution at $130 \mu$m, however, would mean the same grains would be too efficient at producing scattering polarization at Bands 6 and 7.

% convolution of rings and gaps? gaps have more obvious radiation anisotropy effects
%HL Tau is known for its prominent substructure with multiple rings and gaps. The low resolution polarization data 
% dust settling and seeing different layers 
%% spherical grains is not a good assumption. Spheroidal grains also suffer from resonance. Irregular grains could alleviate the tension
% Spherical grains may not be a good assumption 
% the refractive index across wavelength is completely off? 
The fact that we do not find a combination of disk and dust parameters that satisfies the empirically derived constraints on the scattering component $S$ could signify that our adopted dust model is drastically unrepresentative of the true properties of grains in disks. The dust mixture and its resulting refractive index remain poorly understood \citep[e.g.][]{Birnstiel2018}. For example, \cite{Yang2020} demonstrated that adopting amorphous carbonaceous grains could allow large millimeter grains to produce the observed level of polarization which opens up a larger parameter space for a possible solution. Furthermore, the assumption of spherical grains or ellipsoids introduces severe oscillations in the scattering matrix when the size parameter becomes larger than of order unity. Large irregular grains can produce entirely different scattering behavior compared to large compact spherical grains \citep{Tazaki2019, Munoz2021}. More work is needed to explore the implications of our results on the dust properties.

One caveat is that we have assumed a vertically uniform grain size distribution. While this is plausible given how thin the layer of (sub)millimeter emitting dust is in the HL disk \citep{Pinte2016}, it is not guaranteed. Differential settling of grains of different sizes may provide a solution to the above conundrum \citep{Brunngraber2020, Ueda2021}. Specifically, because of a lower opacity, emission at a longer wavelength comes from closer to the disk midplane, where the grains could be larger. It is conceivable that an increase in the sizes of the grains emitting at a longer observing wavelength may satisfy the constraints shown in Fig.~\ref{fig:sph_S}. For example, at a surface density of $0.1$~g cm$^{-2}$, the constraints are satisfied if the grains emitting at Bands 7, 6, and 3 have maximum sizes of $\sim50$, 100, and 225 $\mu$m respectively. 

Another caveat is that the empirically derived scattering component $S$ is assumed to be independent of the azimuthal location at a given distance from the central star (see Eq. \ref{eq:q_minor} and \ref{eq:q_major}). This is true when the dust at the distance under consideration is optically thick to the photons traveling along the disk plane so that the local radiation field is more or less isotropic. The assumption is reasonable for a geometrically thin dust layer, as appears to be the case for HL Tau \citep{Pinte2016}. If a significant radiation anisotropy in the disk plane exists (e.g., in the radial direction), the equations used to derive the scattering component $S$ need to be modified. 

If there is a significant anisotropy in the radiation field in the radial direction, the following qualitative effects are expected. Photons traveling radially outward along the major axis of the disk are scattered by $90^{\circ}$ to the observer gaining maximal polarization, while those along the minor axis of the disk are scattered by $90^{\circ} \pm i$ to become less polarized. Thus, we expect polarization to be larger along the major axis than minor axis which is opposite to thermal polarization of azimuthally aligned prolate grains. As a result, there could be a net increase in the magnitude of the azimuthal variation of the polarization and the absolute level of $p_{l}$ should be more uniform. The derived thermal component $T_0$, which does not account for the radiation anisotropy, is likely a lower limit to the true level of thermal polarization, because a larger thermal polarization would be needed to compensate for the (opposite) azimuthal variation induced by the scattering of a radially anisotropic radiation field. We will leave a detailed exploration of the quantitative effects of the potential radiation anisotropy to a future investigation.

%% perhaps include some comments on potential monte carlo codes that are considering aligned grains with scattering and motivate the urgency and this paper can serve as a test of consistency: RADMC-3D, polaris, skirt
%As far as we know, there are no Monte Carlo radiation transfer codes that can incorporate scattering of aligned grains. 

% Indeed, the Band 7 image appears similar to a disk without the elliptical thermal polarization and the azimuthal variation could be provided by the radiation anisotropy instead. However, the radiation anisotropy alone cannot explain HL Tau simple because the polarization spectrum will not work.

\subsection{Predictions for Other Wavelengths}
Based on Fig. \ref{fig:q_spectrum}, we can make rough predictions for polarization at other wavelengths. The remaining ALMA bands that can provide continuum polarization are Bands 1, 4, and 5. For Bands 4 and 5 (or $\lambda=1.5$ and $2.1$ mm respectively), we expect that the the scattering polarization $S$ should be in between $0.5\%$ to $0.7\%$ unless either of the bands happen to form a peak in polarization when the optical depth is of order unity (Fig. \ref{fig:q_opt}). We can expect that the thermal polarization will be greater than that at Band 6 and less than that of Band 3. Furthermore, we expect the polarization images of Band 4 and 5 to appear similar to a transition between Bands 3 and 6. Preliminary Bands 4 and 5 data suggests that the predictions seem consistent with the observations (private communication). 

%% Talk about Band 1? it would be really important? what about VLA? 
With the incorporation of Band 1 ($\lambda = 6$ to $8.5$mm) on ALMA, the longer wavelength can potentially probe the HL Tau disk in the optically thin limit thereby avoiding the effects of scattering altogether. Based on previous works \citep{Yang2019, Mori2021} and our results, one can conclude that the grains responsible for the thermal polarization are most likely azimuthally aligned prolate grains in the HL Tau disk. However, it is still important to determine the grain alignment directions with as little contamination from scattering as possible. The polarization image at an optically thinner wavelength should better reveal the intrinsic alignment field which can serve as a more robust test for azimuthally aligned prolate grains. 

Since the optical depth at Band 1 should be lower than that at Band 3, the scattering polarization $S$ should be less than the $\sim 0.5 \%$ at Band 3 and the thermal polarization should increase. Since the Band 3 image is similar to the case of $\tau_{0}=0.05$ in Fig. \ref{fig:model_tau0_conv}, we expect that the polarization image should appear similar to the case of $\tau_{0}=0.01$ with a more obvious contrast between the higher polarization fraction region along the minor axis and the lower polarization fraction region along the major axis (Fig.~\ref{fig:model_tau0_conv}m) and, more strikingly, a vertical bar-like morphology for the polarized intensity after beam convolution (Fig.~\ref{fig:model_tau0_conv}a).

\subsection{Identifying Spiral Alignment}
% Stokes U should be zero along the principle axes
% show how the Stokes U would shift given some spiral alignment
% it could be due to convolution of an elongated beam 
% High angular resolution data would be beneficial 
\begin{figure*}
    \centering
    \includegraphics[width=\textwidth]{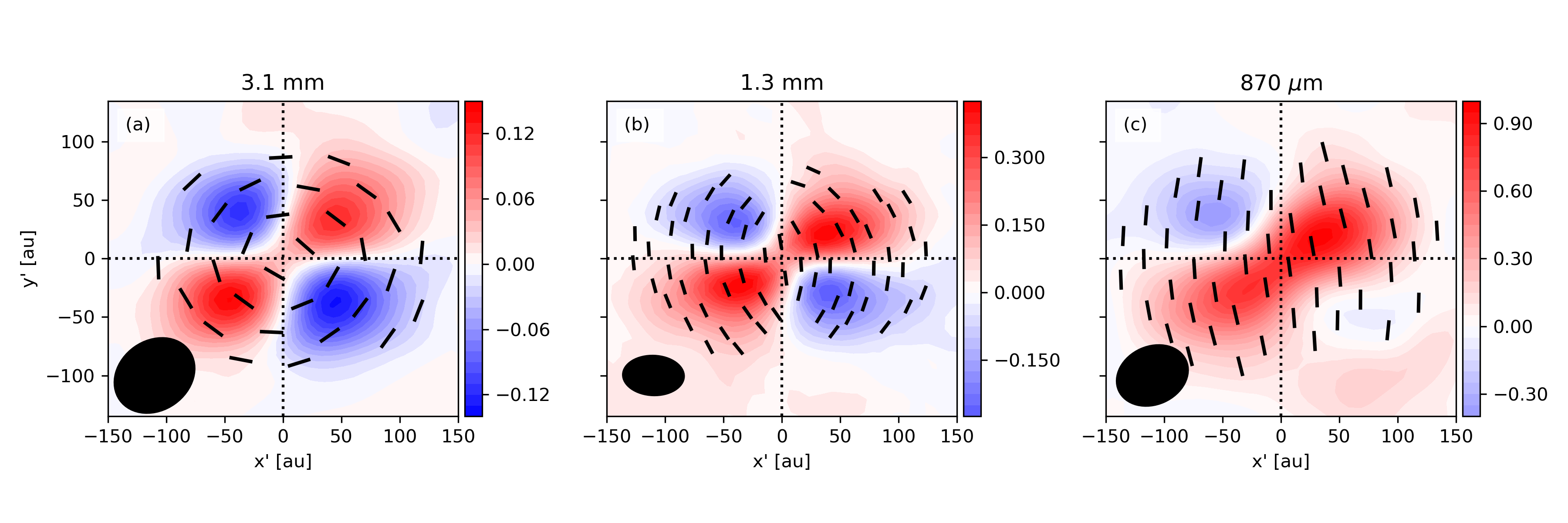}
    \caption{
        The Stokes $U'$ images in mJy beam$^{-1}$ of HL Tau at each wavelength. The black dotted lines correspond to $x'=0$ and $y'=0$. The vectors show the linear polarization angle. The black ellipse is the beam resolution. A disk with perfect azimuthally aligned prolate grains should have Stokes $U'=0$ along the major and minor axes. 
    }
    \label{fig:stokesU}
\end{figure*}

From Section \ref{sec:hl_tau}, we have showed that scattering mainly provides a positive Stokes $Q'$ in the principle frame. The Stokes $U'$ on the other hand, is not heavily affected by scattering (Fig. \ref{fig:qu_azi}) and retains the underlying thermal polarization (see also Section~\ref{ssec:azimuthal}). A natural result is that along the major and minor axis, we should have Stokes $U'=0$ if the prolate grains are perfectly aligned in the toroidal direction. 

Fig.~\ref{fig:stokesU} shows the Stokes $U'$ for Bands 3, 6, and 7 without folding. We further plot the lines, $x'=0$ and $y'=0$, which is where Stokes $U'$ should equal zero for an axisymmetric disk. Along the minor axis of the Band 3 Stokes $U'$, the path where Stokes $U'=0$ has a slight clock-wise tilt from $x'=0$ and less of a tilt from $y'=0$. For Band 6, Stokes $U'=0$ also has a slight clock-wise tilt from $y'=0$. These deviations can be due to an elongated beam that is not parallel to the major or minor axis.\footnote{The offset could also be due to uncertainty of the position angle of the disk major axis in general, but it is unlikely the case for HL Tau given the tight constraint from available high angular resolution images \citep{almapartnership2015}.} However, we demonstrate that it can also be due to a spiral alignment. 

For illustration, we use the reference model in Section \ref{sec:hl_tau}. We let 
\begin{align}
    \phi = 2 \pi - \phi_{d} + \delta 
\end{align}
where $\delta$ is the tilt between the $y$-axis of the slab and the radial direction of the disk in the counterclockwise direction when the disk is viewed face-on. 

%We select difference levels of $\delta$ and show the polarization images in Fig. \ref{fig:model_spiral}. The left column shows the case of $\delta=0^{\circ}$, which corresponds to perfect toroidal alignment (the left column of Fig. \ref{fig:model_spiral} corresponds to the right column of Fig. \ref{fig:model_tau0_conv}).

\begin{figure*}
    \centering
    \includegraphics[width=\textwidth]{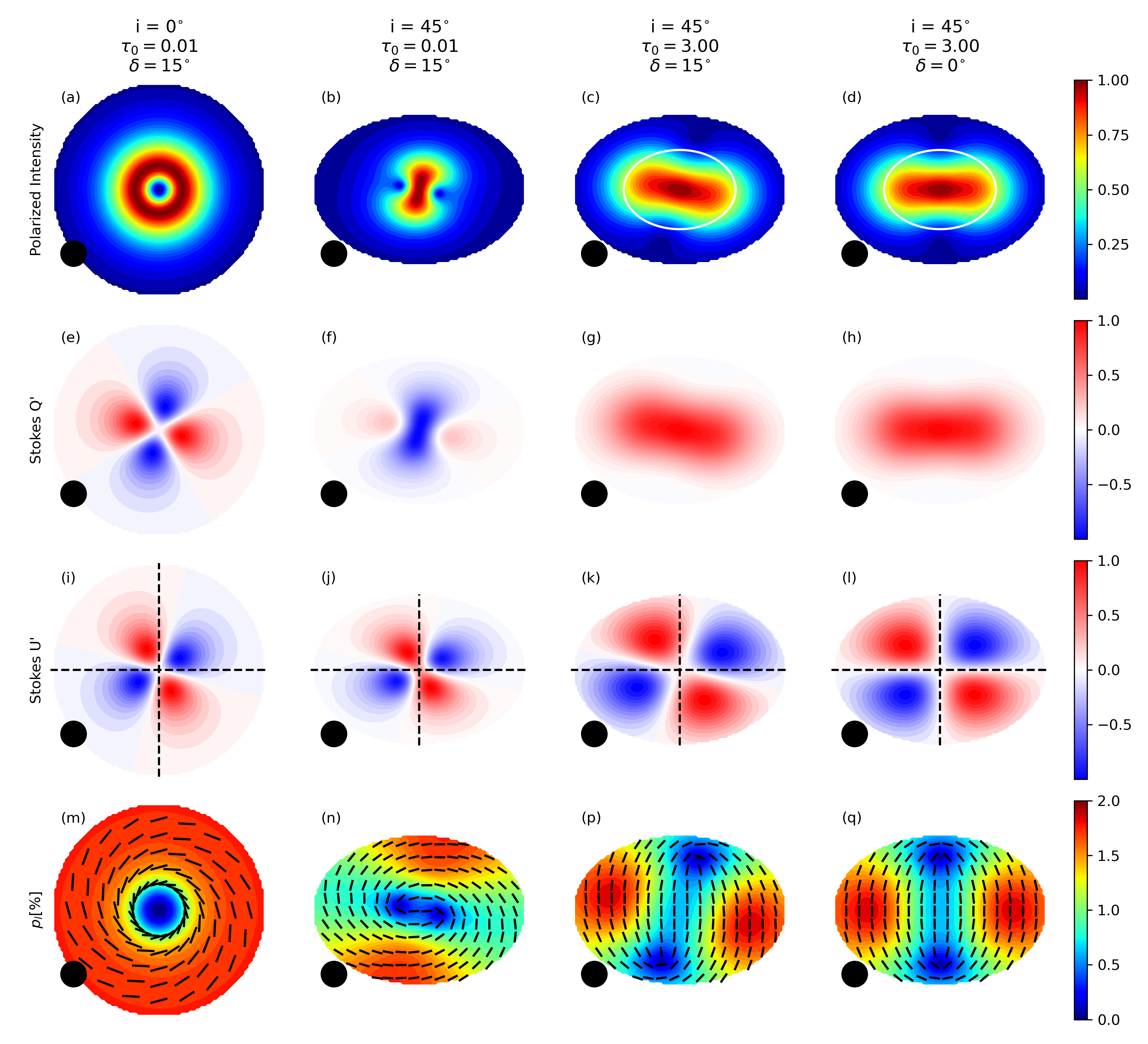}
    \caption{
        The polarization images to illustrate effects of spiral alignment. The polarization properties are plotted in the same way as Fig. \ref{fig:obs_rot}: the top row is the polarized intensity relative to its peak; the second and third row are the Stokes $Q'$ and $U'$ relative to its peak; the last row is the polarization percent and the vectors are the polarization direction. The white contour in the top row is when the optical depth is unity. The black dashed lines in the third row marks $x'=0$ and $y'=0$. The black circle in the bottom left is the beam size. The leftmost column is a disk with spirally aligned grains when optically thin ($\tau_{0}=0.01$) and viewed face-on. The second column is the same optically thin case but inclined at $i=45^{\circ}$. The third column is when the optical depth increases ($\tau_{0}=3.0$). The rightmost column is when the grains are perfectly azimuthally aligned (corresponds to the rightmost column of Fig.~\ref{fig:model_tau0_conv}). 
    }
    \label{fig:model_spiral}
\end{figure*}

%% first explain the physical reason
In the optically thin and face-on case (left column of Fig.~\ref{fig:model_spiral}), the polarization angles directly trace the long axis of the grain which is tilted away from the perfect azimuthal pattern when $\delta=15^{\circ}$ (Fig.~\ref{fig:model_spiral}m). The polarization fraction in the center is low because of beam averaging. The Stokes $Q'$ and $U'$ is rotated in the clock-wise direction as seen in the plane-of-sky (Fig.~\ref{fig:model_spiral}e,i). The rotation can be easily understood by the following. Consider a grain at some positive $x'$ (left of the vertical axis as defined in Section \ref{sec:hl_tau} or $\phi_{d}=0^{\circ}$) that is azimuthally aligned. The Stokes $Q'$ is positive and Stokes $U'$ should be $0$ because the projected long axis of the grain is in the vertical direction. With the extra spiral angle $\delta=15^{\circ}$, the grain rotates counter-clockwise and results in a positive Stokes $U'$. The rotation applies to each location in the disk and thus Stokes $Q'$ and $U'$ are effectively rotated. 

When the optically thin disk is inclined (second column of Fig.~\ref{fig:model_spiral}), the azimuthal variation in $p_{l}$ appears. Similar to the azimuthally aligned case (left column of Fig.~\ref{fig:model_tau0_conv}), prolate grains along the disk major axis is viewed more pole-on, while those along the disk minor axis is viewed more edge-on. However, the largest polarized intensity and $p_{l}$ are no longer along the disk minor axis, but slightly offset clockwise due to $\delta=15^{\circ}$.

If the optical depth increases, such as $\tau_{0}=3$ of the third column of Fig.~\ref{fig:model_spiral}, scattering polarization dominates and provides a positive Stokes $Q'$ (Fig.~\ref{fig:model_spiral}g). The polarization angles are mostly parallel to the disk minor axis near the center and the effects of spiral alignment is only visible in the outer optically thin regions as seen in Fig.~\ref{fig:model_spiral}p. The regions where Stokes $U'=0$ remains offset from the disk major and minor axes (compare Fig.~\ref{fig:model_spiral}j and k). In contrast, if $\delta=0^{\circ}$ (rightmost column of Fig.~\ref{fig:model_spiral}), the lines where Stokes $U'=0$ is completely along the disk major and minor axes (Fig.~\ref{fig:model_spiral}l).

%% then talk about the simple diagnostic
With $\delta=15^{\circ}$ (third column of Fig.~\ref{fig:model_spiral}), all the polarization properties show modest differences to the perfect azimuthal alignment case (rightmost column of Fig.~\ref{fig:model_spiral}). However, showing $x'=0$ and $y'=0$ as guidelines on Stokes $U'$ makes it easier to identify the deviation. Since scattering does not affect Stokes $U'$ when the projected prolate grain is parallel or perpendicular to $\hat{\theta}$, Stokes $U'$ will be 0 even in the optically thick regions where scattering dominates.

%% what can we learn when applied to HL Tau 
At face value, Stokes $U'$ of Band 3 and 6 both exhibit a slight tilt that appears to suggest $\delta$ much less than the chosen $15^{\circ}$ example in Fig.~\ref{fig:model_spiral} (we find that HL Tau is broadly consistent with $\delta \sim 5^{\circ}$). However, Band 7 does not resemble Bands 3 and 6 and also does not match any of the spiral alignment cases. In particular, the negative Stokes $U'$ in the lower right quadrant almost vanished making it difficult to explain with just spirally aligned grains. Whether or not HL Tau does have spiral alignment will still need numerical confirmation, especially considering the elongated beam, which is beyond the scope of this paper. Higher angular resolution will help the simple diagnostic, since the lines where Stokes $U'=0$ can be better resolved and differentiated from the beam convolution effects. The use of plane-parallel slab models as demonstrated in this section can constrain the alignment direction of the grains under the influence of scattering and may help identify the alignment mechanism of disks.

%% caveat on folding 
Note that folding across the minor axis and major axis described in Section \ref{ssec:fold_image} do not hold for spirally aligned grains. Instead, the image has rotational symmetry of order 2 if the disk is both geometrically thin and axisymmetric. 

\subsection{Implication for Other Sources}

There are other sources that are similar to HL Tau which exhibit the scattering polarization pattern at the shorter wavelength, typically Band 7, and an azimuthal pattern at the longer wavelength, e.g., Band 3. The DG Tau disk observed at Band 3 \citep{Harrison2019} shows a clear azimuthal pattern of polarized intensity with polarization angles that look elliptical. There is a bar of polarized intensity along the major axis with polarization angles parallel to the minor axis of the disk. At Band 7, the polarization angles are largely parallel to the disk minor axis \citep{Bacciotti2018}. The wavelength behavior matches what we expect from an increase of optical depth from Band 3 to 7. The Band 3 polarized intensity is roughly similar to the $\tau_{0}=0.05$ case of the reference convolved model image in Fig.~\ref{fig:model_tau0_conv} with a slightly stronger bar which would suggest a $\tau_{0}$ in between $0.05$ and $0.5$. In contrast, HL Tau does not have a bar at Band 3 which would suggest DG Tau is more optically thick than HL Tau at Band 3. Indeed at Band 7, the polarized intensity peak is offset from the Stokes $I$ peak which is expected if the dust is not settled and the disk is optically thick \citep{Yang2017}. Directly using the major and minor axis polarization fraction at Band 3, we have at $r=0.4\arcsec$, $q_{\text{maj}} \sim 4.5 \%$ and $q_{\text{min}} \sim -6 \%$. Applying Eq. (\ref{eq:q_minor}) and (\ref{eq:q_major}), we get $S\sim0.1 \%$ and $T_{0} \sim 6.1\%$. Band 7 is notably non-settled which makes the technique inapplicable. 

Haro 6-13 is a Class II disk with features similar to DG Tau in its Band 3 polarization image \citep{Harrison2019}. The outer region of the disk exhibits an azimuthal pattern with a bar of polarized intensity along the major axis. In the outer region, the polarization fraction along the minor axis is larger than that along the major axis which is expected from azimuthally aligned prolate grains modulated by little scattering contribution. The Band 7 image only has an obvious polarized intensity near the center with polarization parallel to the disk minor axis (Harrison et al. in prep.). The wavelength behavior is similar to HL Tau and can be explained by scattering of azimuthally aligned prolate grains. 

AS 209 is another Class II source observed at two wavelengths. The Band 7 image has an outer azimuthal pattern with a polarized intensity bar across the major axis \citep{Mori2019}. In the region with an azimuthal pattern, the polarization fraction along the minor axis is just slightly larger than the polarization fraction along the major axis (see Fig.~4 in \citealt{Mori2019}). This would also suggest the presence of azimuthally aligned prolate grains with scattering. However, at the longer wavelength of Band 6, the central polarized intensity bar appears larger \citep{Harrison2021} when we would expect the central bar to diminish and the thermal polarization to grow at the longer wavelength. The cause may be due to different beam sizes. The Band 7 image has a beam of $\sim 0.8 \arcsec$ while the Band 6 images has a beam of $\sim 1.3 \arcsec$. The larger beam of the longer wavelength image may have smeared the stronger polarized intensity due to scattering at the center to the outskirts of the disk where the thermal polarized intensity is low. Higher angular resolution observations of Band 6 will clarify if the contradiction from expectation is due to beam size or a scenario that is not explained by scattering of azimuthally aligned prolate grains like HL Tau.

As pointed out in \cite{Sadavoy2019}, the Band 6 image of the Class I IRS 63 resembles Band 6 of HL Tau in that the polarization in the central region is largely parallel to the disk minor axis suggesting polarization due to scattering, while the outer regions are largely elliptical. Direct comparison with a perfect elliptical pattern shows clear deviations in which the polarization angles have a tendency to follow the direction of the disk minor axis. The deviation is expected if scattering polarization is comparable to the thermal polarization and the extra positive Stokes $Q'$ pushes the polarization angle to follow the disk minor axis (see Fig.~\ref{fig:azi_prof_taum}) as we found with Band 6 of HL Tau.

The accretion disk of the massive protostar, GGD27 MM1, that powers the HH 80-81 jet also has resolved polarized dust continuum at 1.14 mm \citep{Girart2018}. Similar to the previous sources, the polarization fraction shows two distinct regions: the inner region where most of the polarization is parallel to the disk minor axis with $\sim 0.6\%$ polarization and the outer region where the polarization is elliptical with $\sim 6 \%$ polarization. The inner region exhibits a near/far-side asymmetry in the polarized intensity (and polarization fraction) which is expected from scattering if the disk is optically thick and geometrically thick \citep{Yang2017}. In the outer region, the minor axis polarization is larger than the major axis polarization which is expected from azimuthally aligned prolate grains. Given that the disk is likely geometrically thick, our plane-parallel model cannot capture the near/far-side asymmetry nor the radiation anisotropy. Nevertheless, the transition from the inner region to the outer region can also be explained based on a change in optical depth as our Fig.~\ref{fig:model_tau0_conv} demonstrates. 

As a counter example, TMC-1A is a Class I source that cannot be explained by scattering azimuthally aligned prolate grains. The 1.3 mm polarization at the disk center is $\sim 0.7\%$ and parallel to the disk minor axis, while the polarization in the outer region is $\sim 10\%$ and mostly radial \citep{Aso2021}. The radial polarization pattern is better explained by azimuthally aligned oblate grains \citep{Cho2007}, and the central region is most likely due to scattering. At first glance, the decrease to low polarization could be due to dichroic extinction from large optical depth (like the previous cases). However, if the oblate grains are completely aligned azimuthally throughout the disk, we would not expect a depolarized region in between the outer region and the center. Along the disk minor axis, polarization of the outer region from oblate grains share the same polarization angle as that from the central scattering region. Thus, there is no cancellation involved to create the depolarization. Instead, the polarization fraction should transition smoothly (analogous to the $\phi=0^{\circ}$ curve of Fig.~\ref{fig:q_opt}). The reasons behind a depolarized region is likely related to a true change in the grain properties or alignment efficiency. For example, the grains could be less aligned within the depolarized region.

The above discussions lead us to believe that azimuthally aligned scattering grains that appear to explain the multiwavelength polarization of the HL Tau disk may also exist in disks around other low-mass stars and possibly even massive stars. However, not all disks with evidence of aligned grains and scattering can be explained by the same way. Why grains in the HL Tau and other disks are aligned with their long axes along the azimuthal direction is an interesting question that deserves further investigation.

\section{Conclusions} \label{sec:conclusion}
The change of polarization pattern from one wavelength to another in the HL Tau disk has been puzzling since it cannot be explained only by scattering or by aligned grains. In this paper, we offer a consistent treatment of scattering of aligned grains in a plane-parallel slab. Since the (sub)-mm-emitting dust layers in HL Tau and other disks are observed to be geometrically thin, we can use the slab calculations to understand the multiwavelength behavior of HL Tau. Our main results are as follows:
\begin{enumerate}[label=\arabic*)]
    \item The transition of the polarization pattern across wavelength is a natural consequence of a change in optical depth. The increase of scattering at higher optical depth leads directly to the increase of polarization that is parallel to the disk minor axis. At the longest wavelength, Band 3, the azimuthal pattern of the polarization fraction indicates mostly thermal polarization. At a shorter wavelength, like Band 6, the thermal polarization still exists, but it is diminished by dichroic extinction due to the increase in optical depth. At the same time, polarization due to scattering increases also due to the increase in optical depth. At the shortest wavelength, Band 7, the scattering polarization dominates and thermal polarization diminishes further. We identified several sources other than HL Tau that may host similar azimuthally aligned and scattering prolate grains from the multiwavelength transition and the radial transition of polarization. 
    
    \item From the plane-parallel slab model, we find that the polarization from scattering of aligned grains is roughly a linear combination of thermal polarization and scattering polarization from spherical grains if the line of sight is optically thin or moderately optically thick. Scattering polarization from purely spherical grains is constant across the disk azimuth, while the thermal polarization varies along azimuth because the grain is viewed from the edge along the disk minor axis and viewed slightly away from the axis of symmetry along the disk major axis. We devise a simple method to decompose the spatially resolved polarization observed in a disk into a thermal component and a scattering component, based on the fact that the two components add at locations along the major axis and subtract along the minor axis (see Eq. (\ref{eq:q_minor}) and (\ref{eq:q_major})). This empirical decomposition relies on geometric considerations rather than detailed knowledge of the disk or dust properties and, as such, should be relatively robust.  
    
    \item The polarization spectrum supports the idea of scattering aligned grains. We find that the level of polarization from scattering is roughly the same for Band 6 and 7, but decreases at Band 3. The contribution of thermal polarization is the least at Band 7 and gradually increases to Band 3. The behavior of scattering is expected because the polarization from scattering is a constant when the line of sight is optically thick and decreases with optical depth when optically thin. The monotonic increase in thermal polarization to longer wavelength is also expected because the optical depth decreases which leads to less dichroic extinction. However, with a simple (DSHARP) dust model, we cannot find a parameter space in surface density and grain size that can simultaneously explain all three wavelengths of HL Tau. 
    
    \item Rotating the image such that the disk major axis and minor axes are the horizontal and vertical axes of the image (we call the ``principle frame'') aids the interpretation of polarization images. Inclination-induced polarization from scattering only contributes to Stokes $Q$ of the principle frame (expressed as Stokes $Q'$). Directly using the Stokes $U$ of the principle frame (expressed as Stokes $U'$) can serve as a simple way to identify spirally aligned grains even with scattering. For perfect azimuthally aligned prolate grains, Stokes $U'=0$ along the major and minor axis of the disk. With a spiral alignment, the projected prolate grain is no longer parallel to the horizontal or vertical axes of the image which gives a non-zero Stokes $U'$. Instead, the location where Stokes $U'=0$ is away from the major and minor axes. Since scattering mainly contributes Stokes $Q'$ only, spirally aligned grains can be identified from just Stokes $U'$. 
    
\end{enumerate}

\section*{Acknowledgements}

We thank the referee for constructive comments. Z-YDL acknowledges support from the Jefferson Scholars Foundation and also support from the ALMA Student Observing Support SOSPA7-001. ZYL is supported in part by NASA 80NSSC18K1095 and NSF AST-1910106. LWL acknowledges support from NSF AST-1910364. This paper makes use of the following ALMA data: ADS/JAO.ALMA\#2016.1.00115.S and ADS/JAO.ALMA\#2016.1.00162.S. ALMA is a partnership of ESO (representing its member states), NSF (USA) and NINS (Japan), together with NRC (Canada), MOST and ASIAA (Taiwan), and KASI (Republic of Korea), in cooperation with the Republic of Chile. The Joint ALMA Observatory is operated by ESO, AUI/NRAO, and NAOJ. The National Radio Astronomy Observatory is a facility of the National Science Foundation operated under cooperative agreement by Associated Universities, Inc.

% HL Tau: 2016.1.00115.S and 2016.1.00162.S

%%%%%%%%%%%%%%%%%%%%%%%%%%%%%%%%%%%%%%%%%%%%%%%%%%
\section*{Data Availability}

The ALMA data can also be obtained from the ALMA Science Data Archive (\url{https://almascience.nrao.edu/asax/}). Additional data underlying this article are available from the corresponding author upon request.



\bibliographystyle{mnras}
\bibliography{main} % if your bibtex file is called example.bib

%%%%%%%%%%%%%%%%%%%%%%%%%%%%%%%%%%%%%%%%%%%%%%%%%%

%%%%%%%%%%%%%%%%% APPENDICES %%%%%%%%%%%%%%%%%%%%%

\appendix

\section{Consistency Check with Isotropic Scattering} \label{sec:consistency_check}
Under the assumption of isotropic scattering, one can calculate an analytical solution to the emergent Stokes $I$ of the plane-parallel slab (see \citealt{Rybicki1979}, \citealt{Miyake1993}, \citealt{Birnstiel2018}, \citealt{Zhu2019}). Here we use the slab model with spherical grains in Section \ref{sec:slab_results} for simplicity and compare the Stokes $I$ with the analytical solution using the same albedo and optical depth. 

\begin{figure}
    \centering
    \includegraphics[width=\columnwidth]{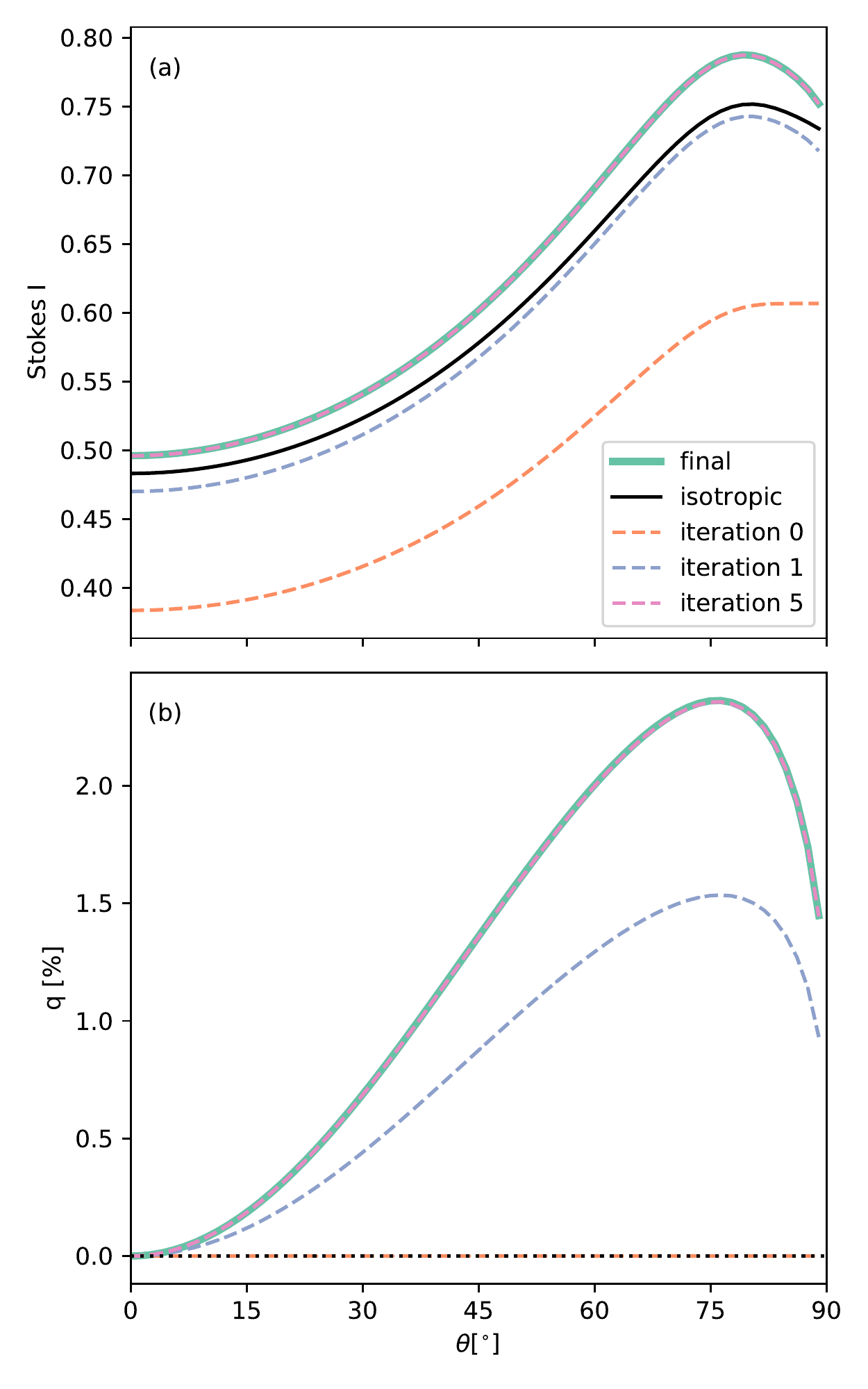}
    \caption{
        The emergent intensity as a function of polar angle $\theta$ for the slab of spherical grains compared to the analytical solution of isotropic scattering. Panel (a) is the Stokes $I$ in units of the black body radiation. Panel (b) is $q$ in percent. For both panels, the solid green lines are the converged solution from the plane-parallel slab. The dashed colored lines are the results from the 0th, 1st, and the 5th iteration. For panel (a), the black solid line is the solution from the analytical solution of isotropic scattering. The black horizontal dotted line in panel (b) marks $q=0$. Note that the curves from ``iteration 5'' is on top of the curves from the final iteration.
    }
    \label{fig:iter_iso}
\end{figure}

In Fig.~\ref{fig:iter_iso}a, the Stokes $I$ from isotropic scattering is similar to the final iteration of Stokes $I$ of the plane-parallel slab. For both analytical and numerical cases, the Stokes $I$ is higher at higher $\theta$, because the optical depth along the line of sight increases with increasing $\theta$. Differences between the two cases are expected, since the analytical solution ignores how scattering depends on the polarization state, but the analytical solution evidently captures most of the Stokes $I$. Note that the analytical solution cannot produce Stokes $Q$ under the assumption of isotropic scattering. 

We also show the emergent intensity of the radiation field on the path of iterating to convergence. The final solution took 13 iterations to converge. At ``iteration 0,'' the emergent intensity is equal to the non-scattering slab case by definition. The resulting $q$ in Fig. \ref{fig:iter_iso}b is thus zero. The ``iteration 1'' case scatters the thermal radiation field once and the Stokes $I$ is already comparable with the isotropic scattering case. With one scatter, the grains can produce polarization as shown in Fig.~\ref{fig:iter_iso}b. With ``iteration 5,'' the Stokes $I$ and $q$ are nearly the same as the final solution.

\section{Converting between Laboratory Frame and the Grain Frame} \label{sec:lab_to_grain_frame}
The Stokes parameters are defined with respect to a reference plane. In the lab frame of Fig.~\ref{fig:coordinates}, the reference plane is the plane formed by $\hat{z}$ and $\hat{n}$. In the grain frame, the reference plane includes the axis of symmetry ($x$-axis) and $\hat{n}$. The viewing angle $\theta_{g}$ in the grain frame is the angle between the $x$-axis and the line of sight. The relation between the Stokes parameters in the lab frame and those in the grain frame involves a rotation of the reference plane. 

Following \cite{Mishchenko2000}, if the reference plane is rotated by an angle $\psi$ in the anticlockwise direction when looking in the direction of propagation, the original Stokes $\vect{I}$ is expressed as Stokes $\vect{I}'$ related by a 4-by-4 rotation matrix $\matr{L}$
\begin{align} \label{eq:stokes_rotation}
    \vect{I}' = \matr{L}(\psi) \vect{I} \equiv \begin{bmatrix}
        1 & 0 & 0 & 0 \\
        0 & \cos 2\psi & \sin 2 \psi & 0 \\
        0 & - \sin 2 \psi & \cos 2 \psi & 0 \\
        0 & 0 & 0 & 1
    \end{bmatrix}
    \vect{I} \text{ .}
\end{align}
For our setup, the Stokes parameters in grain frame $\vect{I}_{g}$ is related to $\vect{I}$ by $\vect{I}_{g} = \matr{L}(-\psi) \vect{I}$. 

We can solve for $\theta_{g}$ and $\psi$ from the following. Let the origin be $O$ and let $A$, $B$, $C$ be the points on the unit sphere intersected by the $\hat{x}$, $\hat{z}$, and $\hat{n}$. From the spherical triangle $ABC$, we first have
\begin{align}
    \cos \theta_{g} = \sin \theta \cos \phi 
\end{align}
from the spherical law of cosines which gives Eq.~(\ref{eq:theta_g}). $\psi$ is the angle formed by arcs $CA$ and $CB$, and we easily find that 
\begin{align}
    \cos \psi &= - \dfrac{ \cos \theta \cos \theta_{g} }{ \sin \theta \sin \theta_{g} } \\
    \sin \psi &= \dfrac{ \sin \phi }{ \sin \theta_{g} }
\end{align}
through the spherical law of cosines and sines. Reorganizing and applying \ref{eq:stokes_rotation} will give the Stokes parameters in the slab. Following the conversions of the slab to the disk in Section \ref{sec:hl_tau}, i.e., $i=\theta$, Eq.~(\ref{eq:phi_to_phi_d}), and a sign change of Stokes $U$, we can obtain the contribution of thermal polarization to Eq.~(\ref{eq:rotated_analytical_q}) and (\ref{eq:rotated_analytical_u}).

\section{Convergence Study on The Number of Grid Points } \label{sec:convergence_gridpoints}

For all the models presented in the main text, we have adopted $N_{\mu}=32$, $N_{\phi}=32$, and $N_{z}=128$ (Section~\ref{sec:plane_parallel}). To test if the number of grid points, $N_{\mu}$, $N_{\phi}$, and $N_{z}$ is enough, we compare the emergent Stokes $I$, $Q$, and $U$ by varying the number of grid points. As a reference, we set $\lambda=1$mm, $x=0.4$, $s=0.975$, and $\tau_{m}=3$.

\begin{figure*}
    \centering
    \includegraphics[width=\textwidth]{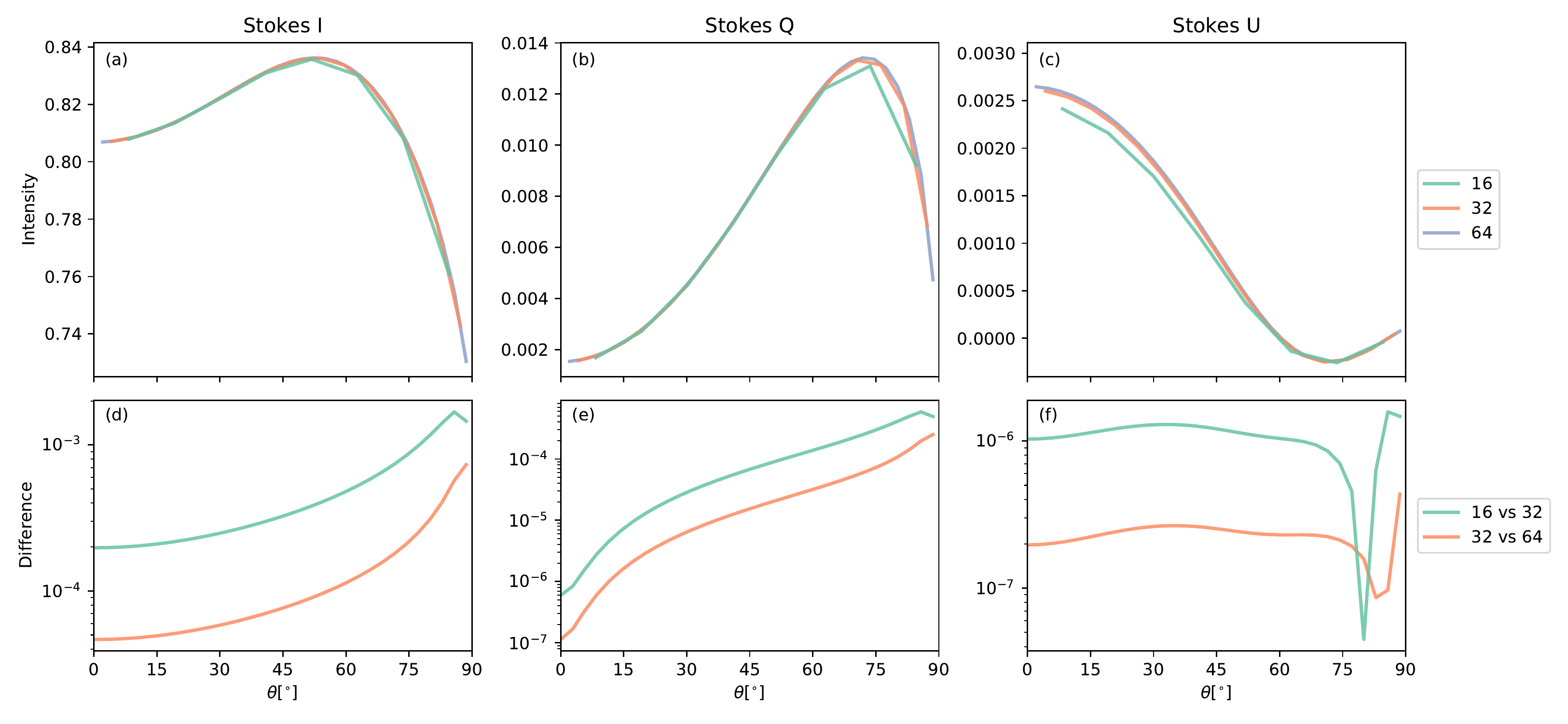}
    \caption{
        Top row: The emergent $I$, $Q$, and $U$ (in units of the thermal radiation of the slab) as a function of $\theta$ at $\phi=30^{\circ}$. The legend corresponds to the adopted number of $N_{\mu}$ and $N_{\phi}$. Bottom row: The absolute difference between consecutive grid resolutions.
        }
    \label{fig:theta_gridding_ster}
\end{figure*}

First, we compare cases with different resolutions in the steradian angles. We fix $N_{z}=128$ and consider three different cases with $N_{\mu}=N_{\phi}=16$, $32$, and $64$. The top row of Fig.~\ref{fig:theta_gridding_ster} shows the Stokes $I$, $Q$, and $U$ profiles as a function of $\theta$ for $\phi=30^{\circ}$. Instead of post-processing the emergent intensity (as was done in Section~\ref{sec:plane_parallel}), we show the emergent intensity as derived on the native grid points to visualize the effects of increasing the angular grid points. For a quantitative comparison, we post-process the emergent intensity onto a common $\theta$ grid and calculate the absolute difference (bottom row of Fig.~\ref{fig:theta_gridding_ster}). As a reference, the relative difference between 32 and 64 grid points for $I$, $Q$, $U$ at $\theta=45^{\circ}$ is $\sim0.009\%$, $\sim 0.2\%$, and $\sim 0.03\%$ respectively.

\begin{figure*}
    \centering
    \includegraphics[width=\textwidth]{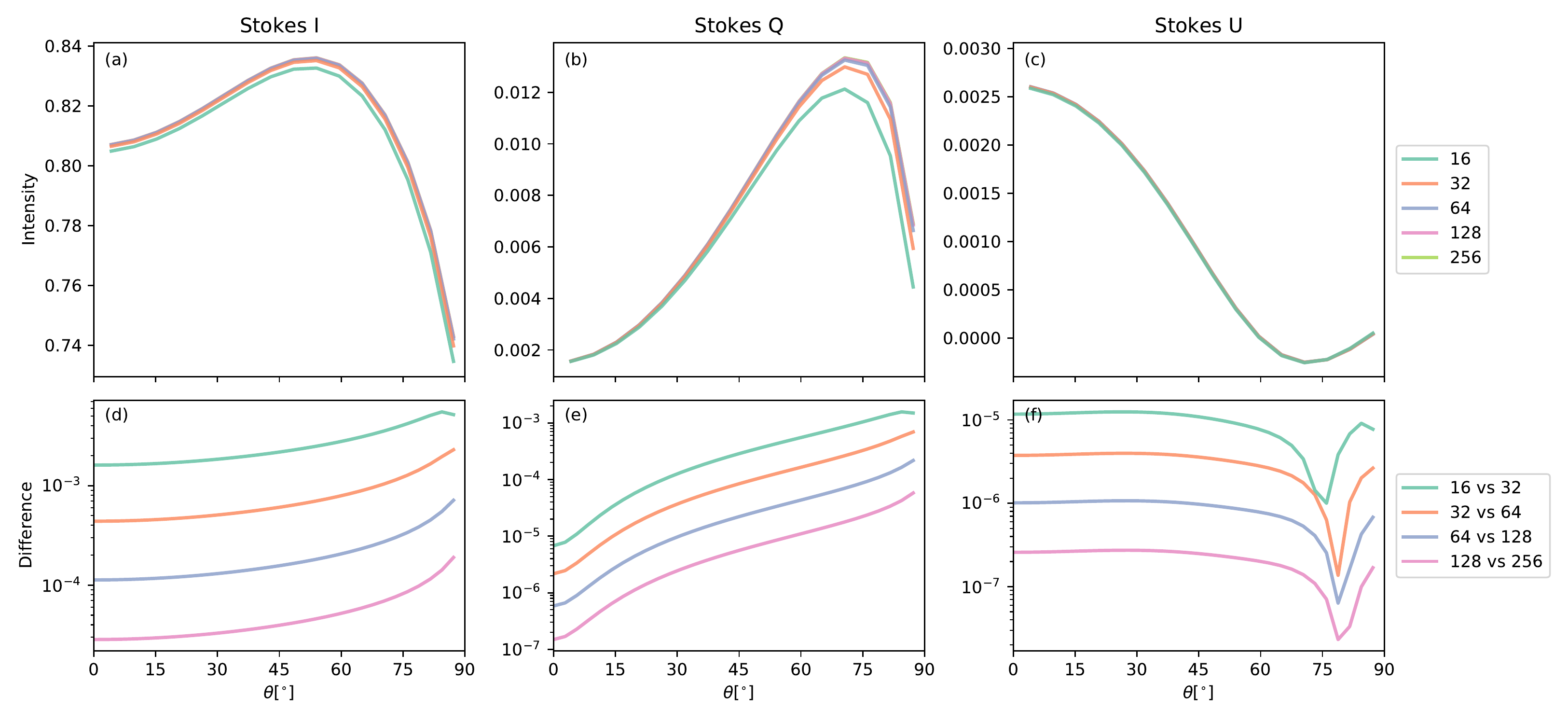}
    \caption{
        Same as Fig.~\ref{fig:theta_gridding_ster}, but comparing different $N_{z}$. 
        }
    \label{fig:theta_gridding_z}
\end{figure*}

Fig.~\ref{fig:theta_gridding_z} shows the comparison of cases with different values of $N_{z}$. We fix $N_{\mu}=N_{\phi}=32$ and consider $N_{z}=16$, $32$, $64$, $128$, and $256$. As with the previous case, we show the solved intensity in the top row and then show the difference between different $N_{z}$ in the bottom row. 
As a reference, the relative difference between 128 and 256 grid points for $I$, $Q$, and $U$ at $\theta=45^{\circ}$ is $\sim0.005\%$, $\sim0.07\%$, and $\sim 0.03\%$ respectively.

\section{Effects of More Elongated Prolate Grain} \label{sec:elongated_prolate}

In this appendix, we show the azimuthal variation of a slab with a more elongated prolate grain. We adopt the same slab presented in Section~\ref{sec:slab_results} ($x=0.4$, $\lambda=1$mm), but set $s=0.5$.  Fig.~\ref{fig:azi_prof_taum} shows the azimuthal variation of $q$, $u$, $p_{l}$, and $\zeta$ at $\theta=45^{\circ}$ for different values of $\tau_{m}/\mu$. 

\begin{figure*}
    \centering
    \includegraphics[width=\textwidth]{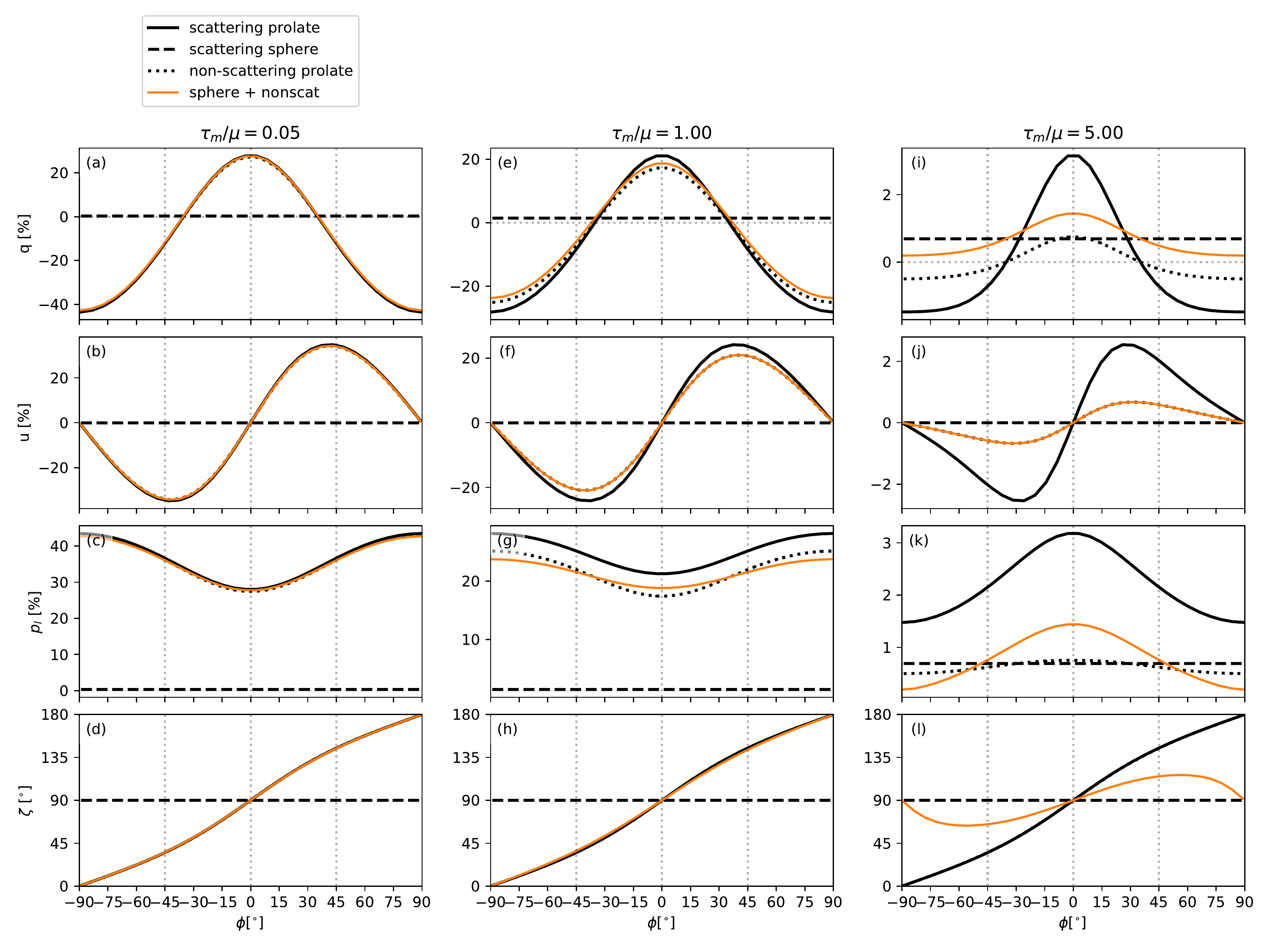}
    \caption{
        Same as Fig.~\ref{fig:azi_prof_taum} but for $s=0.5$.  
    }
    \label{fig:azi_prof_taum_s0.5}
\end{figure*}

Compared to Fig.~\ref{fig:azi_prof_taum}a, Fig.~\ref{fig:azi_prof_taum_s0.5}a shows a distribution of $q$ of the scattering prolate grains that has a larger variation (up to $|q|\sim 40\%$ at $\phi=\pm 90^{\circ}$) because the prolate grain is more elongated which emits a higher thermal polarization. The large level of polarization is not too surprising, since the dipole approximation, using Eq.$\ref{eq:p0_dipole}$ with $s=0.5$, gives $p_{0}\sim43\%$. The level of variation of $q$ decreases for $\tau_{m}/\mu=1$ (Fig.~\ref{fig:azi_prof_taum_s0.5}e) because of the increase in optical depth (and thus dichroic extinction). However, in contrast to Fig.~\ref{fig:azi_prof_taum}e, the level of thermal polarization (shown as dotted lines) remains comparable to the $q$ of the scattering prolate grains, while the $q$ from spherical grains is vastly incomparable. 

In the most optically thick case, $\tau_{m}/\mu=5$, the $q$ of scattering prolate grains with $s=0.5$ (Fig.~\ref{fig:azi_prof_taum_s0.5}i) is very different from the $q$ of grains with $s=0.975$ (Fig.~\ref{fig:azi_prof_taum}i). In both cases, scattering dominates the polarization, but the case with $s=0.975$ produces a near constant level of $q$, while the case with $s=0.5$ produces $q$ that is positive at $\phi=0^{\circ}$ and negative at $\phi=\pm 90^{\circ}$. This is because the elongated grain has a drastically different scattering matrix compared to the spherical grain case \citep[e.g.][]{Yang2016_oblate, Kirchschlager2020}. We conclude that, in this case of highly elongated grains, the simple linear decomposition model, which used spherical grains to approximate contributions from scattering, does not resemble the complete treatment of scattering prolate grains.

Given that the maximum $p_{l}$ at Band 3 of HL Tau is $\sim 2\%$ and the Band 7 polarization is unidirectional (as opposed to switching signs like in the optically thick case with $s=0.5$), the prolate grains cannot be too elongated. Therefore, it is reasonable to assume that the linear decomposition applies. However, this is under the assumption of perfect alignment. Imperfect alignment of highly elongated grains is an interesting possibility that remains to be addressed, but it is beyond the scope of this paper.

\section{Effects of Larger Size Parameter} \label{sec:larger_prolate_azimuthal}

In Section~\ref{sec:slab_results}, we adopted a size parameter of $x=0.4$ which gave an albedo $\sim 0.4$ that made contributions from scattering and thermal emission comparable. In this appendix, we explore the effects of a larger contribution from scattering by increasing the size parameter. This is motivated by \cite{Carrasco2019} which inferred an albedo of $\sim 0.9$ at millimeter wavelength. We use the same slab presented in Section~\ref{sec:slab_results} ($s=0.975$, $\lambda=1$mm), but change $x$ to $1$ to get an albedo of $\sim 0.9$.

\begin{figure*}
    \centering
    \includegraphics[width=\textwidth]{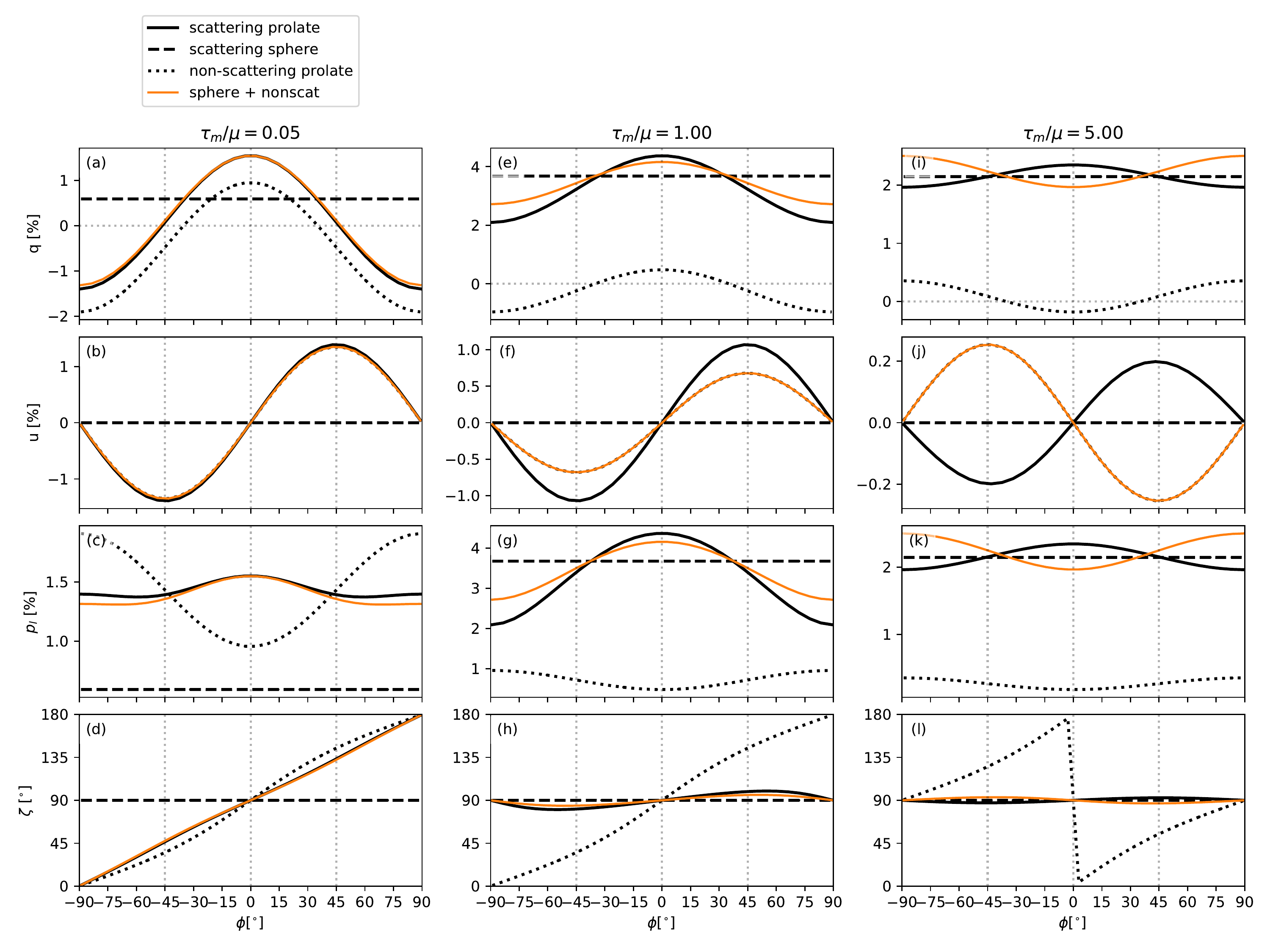}
    \caption{Same as Fig.~\ref{fig:azi_prof_taum} but for $x=1$.}
    \label{fig:azi_prof_taum_x1.0}
\end{figure*}

Fig.~\ref{fig:azi_prof_taum_x1.0} shows the azimuthal profiles of $q$, $u$, $p_{l}$, and $\zeta$ for $\tau_{m}/\mu=0.05$, $1$, and $5$ plotted in a similar fashion as Fig.~\ref{fig:azi_prof_taum}. For $\tau_{m}/\mu=0.05$, which represents the optically thin case, the polarization is still largely dominated by thermal polarization (Fig.~\ref{fig:azi_prof_taum_x1.0}a) similar to when $x=0.4$ (Fig.~\ref{fig:azi_prof_taum}a). However, $p_{l}$ from $x=1$ (Fig.~\ref{fig:azi_prof_taum_x1.0}c) is already qualitatively different $p_{l}$ from $x=0.4$ (Fig.~\ref{fig:azi_prof_taum}c) in that the former case peaks at $\phi=0^{\circ}$ while the latter case peaks at $\phi=\pm90^{\circ}$. Evidently, the larger albedo allows a slightly larger contribution of polarization from scattering even for $\tau_{m}/\mu=0.05$ which evens out the azimuthal variation of $p_{l}$ (recall the addition and canceling effects at different azimuth from Section~\ref{ssec:azimuthal}).

When $\tau_{m}/\mu=1$, $q$ of scattering aligned grains ranges from $2\%$ to $\sim 4\%$ (Fig.~\ref{fig:azi_prof_taum_x1.0}e) which is larger than $q$ of $x=0.4$ (Fig.~\ref{fig:azi_prof_taum_x1.0}e). The large boost in $q$ is due to the larger contribution from scattering which is to be expected since $q$ of scattering $x=1$ spherical grains is at a much higher level as well (Fig.~\ref{fig:azi_prof_taum_x1.0}e). When $\tau_{m}/\mu=5$, $q$ of scattering aligned grains drops to $\sim 2\%$ and the effects of thermal polarization is minimal (Fig.~\ref{fig:azi_prof_taum_x1.0}i). Note that $q$ from non-scattering prolate grain has reversed signs compared to its optically thin counterpart, because dichroic extinction takes out more polarization than what its thermal polarization can produce (see Section~\ref{ssec:optical_depth}).

We can identify that when the albedo is large, $q$ of scattering aligned grains is also approximated by $q$ from thermal polarization plus $q$ from scattering spherical grains for $\tau_{m}/\mu=0.05$, but the approximation begins to deviate slightly more so than with $x=0.4$ when $\tau_{m}/\mu=1$. This is expected since the scattering plays a larger role even when the slab is moderately optically thick. For $\tau_{m}/\mu=1$, the scattering optical depth for $x=0.4$ is $0.4$, but the scattering optical depth for $x=1$ is $0.9$. In the optically thick case ($\tau_{m}/\mu=5$), the approximation breaks down completely, similar to when $x=0.4$ (see Appendix~\ref{sec:elongated_prolate} for another example). 

With the adopted $x=1$, $p_{l}$ is greater than $\sim 2\%$ when $\tau_{m}/\mu \geq 1$, but the high level of polarization is not seen in HL Tau across the three bands. At face-value, it may appear that the size parameter of the grains should not be greater than unity, similar to conclusions derived from assuming spheres \citep{Yang2016_inc, Kataoka2016_hltau}. However, the level of polarization also depends on size distribution and composition which remains uncertain.

%%%%%%%%%%%%%%%%%%%%%%%%%%%%%%%%%%%%%%%%%%%%%%%%%%

% Don't change these lines
\bsp	% typesetting comment
\label{lastpage}
\end{document}